\documentclass[preprint,authoryear,11pt]{elsarticle}
\usepackage[letterpaper, left=.8in, top=.8in, right=.8in, bottom=.8in]{geometry}
\newcommand{\blind}1
\def\dcmm{\if1\blind BerryWest2018dcmm\else blinddcmmref\fi}
\graphicspath{./}  

\usepackage{color,charter,enumerate,graphicx,subcaption,natbib,here,setspace,multirow,amsmath,amssymb,amsfonts,dsfont,placeins}
\usepackage{booktabs, array}  
\newcolumntype{M}[1]{>{\centering\arraybackslash}m{#1}}
\usepackage[usenames,dvipsnames,x11names]{xcolor}
\usepackage[pdftex,pagebackref=true]{hyperref}
\hypersetup{colorlinks,linkcolor=RoyalBlue1,urlcolor=RoyalBlue2, citecolor=RoyalBlue3}

\def\cD{{\cal D}}\def\cI{{\cal I}} \def\cM{{\cal M}} 
\def\seq#1#2{#1{:}#2}  
\def\eqno#1{eqn.~(\ref{eq:#1})}\def\eqnotwo#1#2{eqns.~(\ref{eq:#1},\ref{eq:#2})}
\def\ind#1{\mathds{1}(#1)} 
\def\F{\mathbf{F}} \def\G{\mathbf{G}} \def\W{\mathbf{W}}  \def\f{\mathbf{f}}
 
 \def\bzero{\mathbf{0}} 
\def\m{\mathbf{m}} \def\C{\mathbf{C}}\def\a{\mathbf{a}} \def\R{\mathbf{R}} \def\H{\mathbf{H}} \def\P{\mathbf{P}}
\def\btheta{\boldsymbol{\theta}}\def\bomega{\boldsymbol{\omega}} \def\bbeta{\boldsymbol{\beta}} \def\bgamma{\boldsymbol{\gamma}}
\def\bphi{\boldsymbol{\phi}}
\def\bxi{\boldsymbol{\xi}}

\begin{document}

\begin{frontmatter}
\title{Probabilistic forecasting of heterogeneous \\ consumer 
			transaction-sales time series}  
\author{
\begin{center}
\begin{tabular}{ccc}
\large Lindsay R. Berry $^a$ & \large Paul Helman $^b$  & Mike West $^a$\\ 
\footnotesize\href{mailto:Lindsay.Berry@duke.edu}{Lindsay.Berry@duke.edu} & 
\footnotesize\href{mailto:Paul.Helman@8451.com}{Paul.Helman@8451.com} &
\footnotesize\href{mailto:Mike.West@duke.edu}{Mike.West@duke.edu}\\
\end{tabular}
\\
\bigskip
\small\em 
$^a$ Department of Statistical Science, Duke University, Durham 27708-0251. U.S.A.\\
$^b$ 84.51$^\circ$, 100 West 5th Street, Cincinnati, OH 45202.  U.S.A.\\
\end{center}
}
\begin{abstract}
We present new Bayesian methodology for consumer sales forecasting. With a focus on multi-step ahead forecasting of daily sales of many supermarket items, we adapt dynamic count mixture models to forecast individual customer transactions, and introduce novel dynamic binary cascade models for predicting counts of items per transaction.  These transactions-sales models can incorporate time-varying trend, seasonal, price, promotion, random effects and other outlet-specific predictors for individual items.  Sequential Bayesian analysis involves fast, parallel filtering on sets of decoupled items and is adaptable across items that may exhibit widely varying characteristics.  A multi-scale approach enables information sharing across items with related patterns over time to improve prediction while maintaining scalability to many items. A motivating case study in many-item, multi-period, multi-step ahead supermarket sales forecasting  provides examples that demonstrate improved forecast accuracy in multiple metrics, and illustrates the benefits of full probabilistic models for  forecast accuracy evaluation and comparison. 
\end{abstract}
\begin{keyword}  Bayesian forecasting; decouple/recouple; dynamic binary cascade;  forecast calibration; intermittent demand; multi-scale forecasting; predicting rare events; sales per transaction; supermarket sales forecasting
\end{keyword}
\end{frontmatter}
\newpage

\section{Introduction}\label{sec:intro}

Recent developments in Bayesian state-space modeling for non-negative integer count time series have shown the ability to improve forecast accuracy across heterogeneous time series in both individual and multivariate studies. Examples in forecasting daily sales of supermarket items-- across multiple items at varying levels and exhibiting diverse patterns over time-- show   promise to improve short- to longer-term forecast accuracy and, importantly, to characterize forecast uncertainties in terms of full probabilistic forecast distributions~\citep\dcmm. These dynamic count mixture models (DCMMs) are open to utilizing multiple forms of dynamic regression, trends and seasonal effects, and allow for time variation in effects as well as unpredictable variation in outcomes.  Integration of information on common patterns across related items using Bayesian decouple/recouple concepts for multivariate dynamic models can provide additional forecast improvements while enabling scalability to many item-level time series.

Automated systems for item-specific forecasting of supermarket sales can exploit aspects of such dynamic modeling approaches. Key desiderata are to define full probabilistic forecast distributions for each of many items at the level of individual stores and departments within stores, with a focus on daily sales forecasting over multiple days ahead at each time point.  The aim is to do this  with a model class that is  flexible enough to be tailored to individual products, so as to 
address the enormous diversity experienced in daily sales across many thousands of supermarket items over large numbers of stores in  supermarket chains. Such models must integrate and account for various levels of seasonality (weekly, monthly, yearly), item-level covariates (price/promotion information, local/store-level and holiday effects),   and otherwise allow for and adapt to unpredictable drifts in levels and variability of sales as they arise.  

Challenges  in daily sales forecasting at the store level begin with  many items that sell sporadically, i.e., the so-called intermittent demand problem generating many days with zero sales for such items~\citep[e.g.][]{croston1972forecasting, SyntetosBoylan2005, Teunter2009, arunraj2015, li2018, willemain2004, seaman2018, kolassa2018}. A full probabilistic model must define time-adaptive, item-specific probabilities of zero/non-zero sales patterns, and forecast accuracy assessment must include relevant metrics for probabilistic predictions. 
A second  challenge is that of potential high variability and extreme values in daily sales of items that do sell more frequently, features that have been addressed using various modified Poisson, negative binomial, jump-process models, and others~\citep[e.g.][]{chen2016,chen2017,snyder2012forecasting,mccabe2005bayesian,yelland2009bayesian,Terui2014}.   These and more recent state-space approaches that incorporate dynamic random effects~\citep\dcmm\ can adequately represent unpredictable excess variation and extremes. The challenge is to go further to dissect the observed heterogeneity of outcomes, i.e., to explain and at least partially predict/anticipate diverse levels of variation in sales.   A third    challenge is that of exploiting
cross-series relationships requiring modeling multivariate series of counts~\citep{Soyer2018}.  Here a key aspect of this is 
\lq\lq borrowing strength'', i.e., linking forecasting models across items to share information about  related patterns-- such as patterns of seasonal variation over the days within weeks, and their variation over time-- that may lead to improved quantification of such patterns  to yield improved probabilistic forecasts at the level of individual items and groups of related items.  

These interests in improving item-level forecasting must be understood in the commercial context:  Even very modest improvements in forecast accuracy for a number of individual items can yield substantial practical impact-- in terms of resulting planning and inventory decisions-- at the within-store department, store and system-wide levels.  Finally, we aim for routine, automated analysis applicable to thousands of items on a daily basis across multiple stores, so model implementations must maximally exploit both theoretical tractability and parallelization. 

Section~\ref{sec:model} defines the context of daily supermarket sales forecasting with a multi-step ahead, probabilistic focus, and the new class of dynamic count models incorporating the novel concept of binary cascades.  This   begins with flexible, dynamic count mixture models-- a.k.a. state-space models for heterogeneous count time series-- to assess and forecast 
supermarket item-level demand in terms of transactions events. Coupled with this, development of our dynamic binary cascade concept involves a class of Bayesian non-parametric models to predict numbers of items sold per transaction (or \lq\lq basket''). This is a  new approach involving novel Bayesian dynamic models that are customizable to diverse levels of sales from sporadic/intermittent to persistent levels.  The final component of this section concerns the integration of cross-series information using the novel multi-scale/multivariate time series approach recently introduced in~\cite\dcmm. We adapt this to forecasting transactions rather than sales; this enables relevant data sharing in forecasting item-level demand, which is then coupled with the new binary cascade approach for sales per transaction.  This decouple/recouple framework maximally exploits 
analytic tractability for sequential learning and forecasting for each individual item and enables information sharing across items while maintaining computational scalability; the resulting computational burden remains linear in the number of items.
The technical appendix summarizes relevant model structure and methodological details, including aspects of the forward/sequential Bayesian filtering analysis of the new models, and the consequent details of simulation-based forecasting for sales outcomes prediction over multi-steps ahead at each time point.  
 
Section~\ref{sec:application} develops and showcases a  series of examples of the application of the new model class in analysis and forecasting of supermarket sales with a number of items evidencing substantially differing features in sales levels and variation over time.    Issues of relevant metrics for forecast assessment, including standard point-forecast measures, probabilistic calibration and coverage, are central to the study of applied relevance and role of statistical models. A main ingredient of this study is to promote a more comprehensive understanding of the practical importance of considering a broader range of forecast accuracy assessment summaries. 
Additional comments in Section~\ref{sec:conc} as well as the supporting technical material in the Appendices conclude the paper.

\section{Context and Models \label{sec:model} }

\subsection{Setting \label{sec:contextgoals} }

The modeling advances in this work capitalize on availability of detailed point of sale data on transactions and sales-per-transaction information on supermarket items. Consider one specific item in a given store. Data are observed daily with day $t$ records of (a) the number of transactions involving this item, i.e., of customers purchasing {\em some number} of the item, and (b) for each transaction, the number of units sold. Many items sell sporadically with no or few transactions per day, and with a high probability of only one unit sold per transaction.  Many other items sell more frequently but again generally at 1 or perhaps 2 units per transaction.  Then other items can sell at higher levels per transaction, though again generally small numbers.  Infrequent bursts of item sales occur, often in the context of known promotions or pricing changes. Some items experience rare events in terms of larger numbers of sales in rare batch purchases. 
 
Standing at the end of day $t,$ the forecasting goal is to predict future sales over the coming period of $k$ days; our applied context requires 2-week forecasts, so $k=14.$   We aim to do this in terms of a full probability forecast distribution for that coming period, and this process is repeated each day.   The new model developed dissects and  models item sales by transaction, with the following notation all indexed by day $t$:
\begin{itemize} \itemsep-3pt
\item $y_t$  is the total number of units sold. 
\item $b_t$ is the number of transactions-- or {\em baskets}-- involving at least one unit sale.   
\item $z_t = \ind{b_t>0}$ where $\ind{\cdot}$ is the indicator function; thus $z_t=0$ implies zero transactions, while $z_t=1$ indicates some transactions.
\item $n_{r,t}$ is the number of transactions with {\em more than} $r$ units, where $r=\seq0d$ 
for some specified (small) positive integer $d.$ By definition, $n_{0,t}\equiv b_t.$    Evidently also,  if $n_{r,t}=0$ for some $r\le d$ then
 $n_{r+1,t}= \cdots = n_{d,t}=0.$  
\item $e_t\ge 0$ is the count of {\em excess sales} from any and all transactions that have more than $d$ items. 
 Evidently,  $e_t=0$ unless $n_{d,t}>0.$ 
\item With the above definitions, it follows that   
\begin{equation}\label{eq:salesfrombaskets} 
 y_t = \begin{cases}  0, & \textrm{if } z_t=0,\\ 
 				\sum_{r=\seq 1d} r (n_{r-1,t}-n_{r,t}) + e_t, & \textrm{if } z_t=1. \end{cases}
\end{equation} 
\end{itemize}  
The new dynamic models for forecasting the $y_t$ series are built from coupled components separately modeling transactions $b_t=n_{0,t}$   and the sequence of values $n_{\seq1d,t}, e_t,$ as now detailed.

\subsection{Transaction Forecasting using Dynamic Count Mixture Models (DCMMs)}  \label{sec:DCMM}

First, we utilize a dynamic count mixture model to represent and forecast the item-specific transaction process $b_t$ over time.   This   class of DCMMs provide a flexible framework for modeling non-negative counts that is customized to dealing with zero counts together with potentially diverse patterns of variation of non-zero counts. Two state-space model components are involved. The first is a dynamic binary/logistic regression model for zero/non-zero transactions; the second is a dynamic, 
shifted Poisson log-linear model for transaction levels conditional on there being some transactions.   Each model component may involve covariates-- such as price and promotion predictors, seasonal effect variables, holiday effects, and so forth-- that may partly explain and hence predict variation over time in transaction outcomes.   An initiating application for the development of DCMMs was in forecasting item sales, and one important aspect of these models is that they naturally integrate time-specific random effects--e.g.,  daily random effects in the supermarket forecasting context. This anticipates and adapts to unpredictable levels of variation in outcomes over and above that explained by the conditional Bernoulli and Poisson dynamic models. In sales forecasting, this is particularly key in dealing with relatively common \lq\lq extra-Poisson'' variation and occasional bursts in sales levels.  

The key point here is to adapt DCMMs to model transactions, not sales.   The heterogeneity and over-dispersion seen in sales data is, in part, due to the compounding effect of varying sizes of transactions per customer throughout the day.  When modeling transactions alone,  this level of complexity and diversity in outcomes is diminished; the opportunity for improved forecasting accuracy at the level of transactions is then clear. 

A DCMM for transaction outcomes $b_t$  is defined by a coupled pair of observation distributions 
in which 
\begin{equation}\label{eq:DCMM} 
z_t \sim Ber(\pi_t) \quad\textrm{and}\quad 
   b_t | z_t = \begin{cases} 0, & \quad \text{ if }  z_t = 0, \\ 1 + x_t, \quad x_t \sim Po(\mu_t), &  \quad \text{ if } z_t = 1,\end{cases} 
\end{equation} 
over all time $t.$  Here $Ber(\pi)$ denotes the Bernoulli distribution with success $(z_t=1)$ probability $\pi,$ while $Po(\mu)$ denotes the Poisson distribution with mean $\mu.$  The parameters $\pi_t$ and $\mu_t$ are time-varying according to binary and Poisson 
dynamic generalized linear models~(DGLMs:~\citealp{west1997book}~chapter~15;~\citealp{Prado2010}~section~4.4), respectively; that is, 
\begin{equation} 
\text{logit}(\pi_t) = \F^0_{t} \bxi_t \quad\textrm{and}\quad  
\log(\mu_t) = \F^{+}_{t} \btheta_t \label{eq:dccmstates}
\end{equation}
with latent state vectors  $\bxi_t$ and $\btheta_t$  and known dynamic regression vectors $\F_t^+$  and $\F_t^0$, in an obvious notation. 
The regression vectors can include different covariates and dummy variables,  and the choices can be customized to item.   Some aspects of variation over time-- in both zero/non-zero transaction probabilities and in the conditional levels of non-zero transactions-- comes through the specification of covariates in the regression vectors. Additional aspects of variation can be captured and adjusted for through time variation in the latent state vectors defining time-varying regression coefficients, in the usual state-space mode.  More technical details of model specification and Bayesian filtering/forecasting analyses are given in~\cite\dcmm\ with relevant summaries in~\ref{app:models} here.

\subsection{Dynamic Binary Cascade Models for Sales-per-Transaction} 

A central modeling and methodological innovation here is a new dynamic binary cascade model (DBCM) that directly addresses the interests in precision in dissecting heterogeneity in sales outcomes by focusing on an hierarchical decomposition of numbers of units per transaction. Many items sell just once per transaction, many others sell at perhaps 2 or 3 items, with higher numbers becoming increasingly rare. The multi-scale formulation of a DBCM is motivated by the reality that  predicting rare events of any kind--  here, larger numbers of units per transaction-- is only and properly addressed using hierarchical sequences of conditional probabilities to define chances of outcomes. 
 
The DCMM defines forecast distributions for transactions $b_t$ into the future, and is used to compute predictive probabilities of transaction outcomes as well as-- critically-- to simulate representative future outcomes.  Given a chosen or simulated/synthetic value of $b_t,$ we then condition to model and forecast the daily sales conditional on that level of transactions using the DBCM defined below. In a Bayesian Monte Carlo analysis, repeatedly simulating many representative values of $b_t$ and then sales coupled to each value defines formal computation from the required predictive distribution of sales.  As we move across Monte Carlo samples, uncertainty about transaction levels is represented, and then the conditional uncertainty about sales per transaction factors in.  

Consider then a given a value of $b_t\equiv n_{0,t}.$   The DBCM defines a probability model for $y_t|b_t.$  First, if $b_t=0$ then sales $y_t=0,$  the trivial case.  Consider now cases when $b_t>0$ and  refer to~\eqno{DCMM} to focus on uncertainty about the resulting sales count $y_t.$    The model is structured as follows: 
\begin{itemize} \itemsep-3pt
\item For each $r=\seq1d,$  denote by $\pi_{r,t}$ the probability that the number of items sold per transaction exceeds $r$ given that it exceeds $r-1$, and assume the numbers of units per transaction are conditionally independent across baskets. 
\item  For any number $r=\seq1d,$  the (increasingly small) probability of more than $r$ sales per basket is then implied 
as $\pi_{1,t}\pi_{2,t}\cdots \pi_{r,t}.$ 

This is a key to the strategy and utility of the binary cascade concept: it models and hence forecasts rare events-- unusually high levels sales for any one transaction-- via a sequence of conditional probabilities, each of which is estimable from the data while their product can be very small.

\item  For each $r=\seq1d,$ the hierarchy of sales levels $n_{r,t}$ then follow a sequence of conditional binomial distributions, namely 
$n_{r,t} | n_{r-1,t} \sim Bin(n_{r-1,t}, \pi_{r,t})$ based on these probabilities.   As we sequence through $r=0,1,\ldots,$  if we experience a level $r$ with $n_{r,t}=0$ this implies, of course, that $n_{j,t}=0$ for all $j\ge r.$ 

\item   The excess sales $e_t$ are computed by summing over possible transactions with more than $d$ sales each. 
If $n_{d,t}=0,$ then $e_t=0.$  If, on the other hand, if $n_{d,t}>0$  then $e_t \ge (d+1)n_{d,t}$.

Given that the probability of more than $d+1$ per basket is generally expected to be quite small, the analysis will be quite robust to the conditional distribution of $e_t$.  Hence we consider two strategies to quantifying the excess. One strategy is to leave the distribution of the excess completely unspecified and simply report the probability of $n_{d,t}>0$ along with the forecast distribution of sales $y_t$ {\em conditional on $n_{d,t}=0$}. A second strategy is to simply use a bootstrap analysis in which a simulated forecast with $n_{d,t}>0$ results in randomly sampling the corresponding forecast excess from the empirical distribution of past observed excess values.   This is further discussed and developed in Sections~\ref{sec:forecasting} and~\ref{sec:modelsandpriorsexcess}, and exemplified in the application. 
 
\end{itemize} 
As with the Bernoulli model for zero/non-zero transactions $z_t,$   we have access to the flexible class of dynamic logistic state-space models for each of the elements of the cascade across levels of sales per transaction.  That is, the conditional model of $n_{r,t}$ has the dynamic binomial logistic form 
\begin{equation}\label{eq:DBCM} 
n_{r,t} | n_{r-1,t} \sim Bin(n_{r-1,t}, \pi_{r,t})  \quad\textrm{where}\quad  
	\textrm{logit}(\pi_{r,t}) = \F^0_{r,t} \bxi_{r,t}  
\end{equation}
with latent state vectors  $\bxi_{r,t}$ and   known dynamic regression vectors $\F_{r,t}^0$  in an obvious extension of the earlier notation. 
The regression vectors can include different covariates and dummy variables for each level $r,$ and can be customized to level.   The $\pi_{r,t}$ may be relatively stable over time,  but impacted by price and promotion effects that increase relative probabilities of higher levels of sales per item, so that such information is candidate for inclusion in regression terms.     As with the transaction events, aspects of variation over time comes through the covariates included, but is also potentially represented  via time variation in the latent state vectors $\bxi_{r,t}$ of time-varying regression coefficients.  Additional details of model specification and Bayesian filtering/forecasting analyses are summarized in~\ref{app:models}.

\subsection{Multi-Step Ahead Forecasting  \label{sec:forecasting}} 

Bayesian forecasting   is based on full predictive distributions. In most applications, it is of interest to use direct/forward simulation of multi-step ahead predictive distributions. Among other things, this allows trivial computation of probabilistic forecast summaries for arbitrary functions of the future data over multiple steps ahead. In transactions and sale forecasting,  generating Monte Carlo samples of synthetic futures over a series of days provides forecast summaries for sales each day, the patterns of variation and dependence day-to-day, and other aspects of applied relevance such as cumulative forecasts over a period of days.    Thus, by
\lq\lq forecast'' we now mean simulation -- i.e., the generation of multiple random samples of transactions and sales outcomes over multiple days, defining \lq\lq synthetic'' futures that can be summarized to compute a range of point forecasts of interest under various utility functions, as well as full probabilistic summaries that formally capture and reflect predictive uncertainties. 

Multi-step forecasting via simulation in dynamic transaction-sales models builds on basic simulations from the sets of DGLMs that define model components.  On any day $t$ looking ahead over the next $k$ days based on current information $\{ \cD_t,\cI_t \},$  the requirement is to  generate a large Monte Carlo sample from the full Bayesian predictive distribution for transactions and sales of the item over days $\seq{t+1}{t+k}.$   Denote by superscript $*$ a single Monte Carlo sample of relevant quantities, referred to as a \lq\lq synthetic'' outcome. 
We generate large Monte Carlo samples of outcomes by independently and repeatedly generating single synthetic outcomes as follows. 

\medskip\noindent{\bf\em Forecast Transactions Indicators: }  Over coming days $j=\seq1k$, 
generate the set of $k$ synthetic transactions/no transactions indicators $z_{t+j}^*$  
from the binary DGLM component of the DCMM transaction model.   This is a representative draw from the current $k-$dimensional predictive distribution of $(z_{\seq{t+1}{t+k}}| \cD_t,\cI_t).$   

Technically, this uses direct compositional sampling applying, at each day into the future,  the forward filtering and updating analysis of the binary DGLMs.  This exploits the representation 
$$p(z_{\seq{t+1}{t+k}}| \cD_t,\cI_t) = p(z_{t+1}|\cD_t,\cI_t) \, p(z_{t+2}| z_{t+1}, \cD_t,\cI_t) \cdots p(z_{t+k}| z_{\seq{t+1}{t+k-1}}, \cD_t,\cI_t).$$
Outcomes are simulated by sequencing through the composition here.  Sample $z_{t+1}^*$ from the first component, simply the 
$1-$step ahead distribution implied in the binary DGLM at time $t$. Condition on this value $z_{t+1}^*$ to update the summary information 
in the DGLM, evolve one day and then predict  $z_{t+2}$ using $p(z_{t+2}| z_{t+1}^*, \cD_t,\cI_t);$ this is again just the $1-$step 
ahead distribution in the binary DGLM moved along one day and conditional on the synthetic value $z_{t+1}^*.$ This is recursively 
applied over the following days up to $k-$steps ahead to produce the   full synthetic path $z_{\seq{t+1}{t+k}}^*.$

\medskip\noindent{\bf\em Forecast Non-Zero Transaction Levels: }   For each day ahead $j$ such that $z_{t+j}^*=1,$   generate number of transactions $n_{0,t+j}^*=b_{t+j}^*$ 
from the shifted Poisson DGLM component of the DCMM transaction model.  This gives a  representative draw from the current conditional predictive distribution of  $(b_{\seq{t+1}{t+k}}| z_{\seq{t+1}{t+k}}^*,\cD_t,\cI_t)$ with the implicit 
zero values implied on days such that $z_{t+j}^*=0.$

Technically, this again uses direct compositional sampling, now based on the forward filtering and updating analysis of the Poisson DGLMs.  The concept and format is just as in the above details for the binary DGLM,  simply differing in the distributional forms involved.

\medskip\noindent{\bf\em Forecast Sales per Transaction: } For each day ahead $j$ for which $z_{t+j}^*=1,$  generate a set of basket sizes $n_{1:d,t}^*$  from the dynamic binary cascade model conditional on the  number of transactions  $n_{0,t+j}^*=b_{t+j}^*$.   This gives a  representative draw from the current conditional predictive distribution of  the full sequence of baskets sizes 
$(n_{1:d,\seq{t+1}{t+k}}| b_{\seq{t+1}{t+k}}^*,z_{\seq{t+1}{t+k}}^*,\cD_t,\cI_t)$ with the implicit 
zero values implied on days such that $z_{t+j}^*=0.$
Technically, this  is done by sequencing through the cascade on each day, generating the number of baskets with a single item, and conditional on that number simulating the number with two items, and so on up to $d$ items.   In cases when the  total number of items simulated with fewer than $d+1$ items in any transaction reaches $b_{t+j}^*,$ the implied synthetic number of items sold is established.  Otherwise, the (generally few) remaining transactions involve more than $d$ items each. If the excess distribution is unspecified, then the DBCM outputs the current synthetic probability of the excess sales event $e_{t+j}\ge (d+1) n_{d,t}^*.$ 

If the excess distribution in the DBCM has been specified, we can proceed by simulating from this excess distribution. One specific excess distribution that fits nicely in the compositional forecasting framework is simulating the excess sales from the empirical excess distribution up to time $t$. For example, prior to time $t$, assume we have observed excess sales-per-transactions of $(d+1, \ldots, D)$ with frequencies $(w_{d+1}, \ldots, w_D)$, where $\sum_{i = d+1}^D w_i$ is the total number of transactions with $n_{d,t} > 0$. Given $n_{d,t+k} > 0$, we can forecast the future excess sales $e_{t+k}$ by sampling $n_{d,t+k}$ values with replacement from $(d+1, \ldots, D)$ with weight proportional to $(w_{d+1}, \ldots, w_D)$. 

As with the transactions simulations above, moving ahead over days involves direct compositional sampling, now based on the forward filtering and updating analysis of the sets of conditional binomial  DGLMs.  The concept and format is just as in the above details for the binary DGLM, simply differing in the distributional forms involved.
 
Uncertainty about the underlying DGLM model components are fully accounted for in forward simulation of each of the state vectors. Critically also, each such synthetic outcome inherently reflects day-to-day dependencies as well as  uncertainties about the underlying DGLM model state vectors; that is, we generate full predictive samples from the joint distribution of the binary, Poisson and binomial latent transactions and sales variables over the $k-$step ahead path.   This means that summary inferences on aggregates and other functions of transactions indicators, transactions levels, basket sizes and sales can be directly deduced by simple numerical summaries of the set of Monte Carlo samples.

\subsection{Cross-Series Linkages and Multi-Scale Extensions \label{sec:multi-scale} }

In forecasting multiple items with potentially related patterns over time, the opportunity to improve forecast accuracy by integrating information across series arises.  Introduced in~\cite\dcmm\ in DCMMs for sales forecasting, an approach using dynamic predictors related to cross-series relationships is relevant to potentially both DCMM and DBCM components of the new transaction-sales models here.  The basic idea is to define one or more factors to be used as common predictors in the dynamic regression models for each item.  This is summarized here in the context of a  single DGLM component for each of a collection of (possibly many) time series. Let $N$ be the number of time series and denote by   $\cM_i$ a DGLM component for series $i.$ In the transactions-sales applications,  this can be any one or each of the component binary, binomial and (shifted) Poisson DGLM components.   One particularly relevant context is to share information about related patterns of daily variation withing the week, i.e., weekly seasonal patterns, in which case the DGLM component $\cM_i$  is the shifted Poisson for non-zero transactions for item $i$ at the daily level.    

A multivariate dynamic factor model incorporating cross-series linkages has state and regression vectors defined by 
\begin{equation} \label{eq:Mi} 
\cM_i: \qquad    \btheta_{i,t} =  \begin{pmatrix} \bgamma_{i,t}\\ \bbeta_{i,t} \end{pmatrix},  \quad
			      \F_{i,t}        =  \begin{pmatrix} \f_{i,t}\\  \bphi_t \end{pmatrix}, \qquad i=\seq1N, 
\end{equation} 
with subvectors of conformable dimensions; the  linear predictor is then $ \lambda_{i,t} = \bgamma_{i,t}'\f_{i,t} +\bbeta_{i,t}'\bphi_t$.
Here $\f_{i,t}$ contains constants and series-specific predictors-- such as item-specific prices and promotions in the sales forecasting context. 
The latent factor vector $\bphi_t$ is common to all series-- such as seasonal or brand effects in the sales forecasting context.  Each series has its own state component $\bbeta_{i,t}$ so that the impacts of   common factors   are series-specific as well as time-varying.  

A separate model  depends on $\bphi_t$ and possibly other factors. Denote this model by $\cM_0.$ Forward sequential analysis of data relevant to $\cM_0$ defines posterior distributions for $\bphi_t$ at any time $t$ that can be used to infer and forecast the $\bphi_t$ process as desired.   These inferences on the common factors are then forwarded to each model $\cM_i$ to use in forecasting the individual series. 
Technically, this is done via direct simulation, so that current and future values $\bphi_\ast$  are simulated from the current posterior and predictive distributions under $\cM_0,$ and then forwarded to each $\cM_i$.  At each simulated value, each single posterior and forecast simulation in $\cM_i$ conditions on one sampled $\bphi_\ast$, so that inferences under $\cM_i$ are then available using the standard computations for individual models.  Critically,  the updates and forecasting computations in each $\cM_i$ are performed separately and in parallel, conditional on values of the common factors $\bphi_\ast$;  this decoupling of series for core computations enables scaling in the number $N$ of items, while maintaining the information sharing across items. 

Model $\cM_0$ can be any external model generating information on common factors.  Key special cases 
relevant to  DCMMs for transactions are referred to as multi-scale models. This is highlighted in cases of collections of items within a store that naturally share common patterns of weekly seasonality based on customer traffic through the store. In such cases, $\bphi_t$ may be a scalar factor representing the current day-of-week based on an external model of traffic. The multi-scale special case arises when using 
aggregate transaction data-- such as the total number of transactions on all products, or on some specific subgroup of products-- to define $\cM_0.$ Each item-level model is then built on the predictions about daily variation from the aggregate model, while the elements  $\bbeta_{i,t}$ provide for item-specific, idiosyncratic deviations from the imputed aggregate values.

\section{Application \label{sec:application} }

\subsection{Data \label{sec:data} }

The goal of our case study is to predict future sales of individual supermarket items $\seq{1}{14}-$days ahead. We compare the forecasting performance of the binary cascade framework to a benchmark model; the latter is a DCMM for daily sales as in~\cite\dcmm. This benchmark meets key desiderata of defining full predictive forecasts, flexibility in modeling diverse patterns in series of counts, incorporation of potentially time-varying dynamic seasonal and regression effects, and adaptability to heterogeneous patterns of  otherwise unpredictable variability. 

The  data set records transaction-level purchases of supermarket items in one store of a major retail chain during the 762 day period from June 1st 2015 to July 1st 2017. Each row in the transaction-level data set represents one consumer's purchase of one or more units of a single item. Items are identified by a unique base universal product code (UPC) in the \lq\lq Dry Noodles and Pasta" category. For each transaction event, the data includes item UPC, the purchase date, the effective price per unit, whether or not the item was purchased on promotion, and the unit sales in the given transaction. The daily transactions count for an item is the number of rows on a given day with the item's UPC; the total daily sales is then the sum of unit sales across transactions. 

 \begin{figure}[p]
	\centering
	\begin{tabular}{cc}
		\includegraphics[width=.5\textwidth]{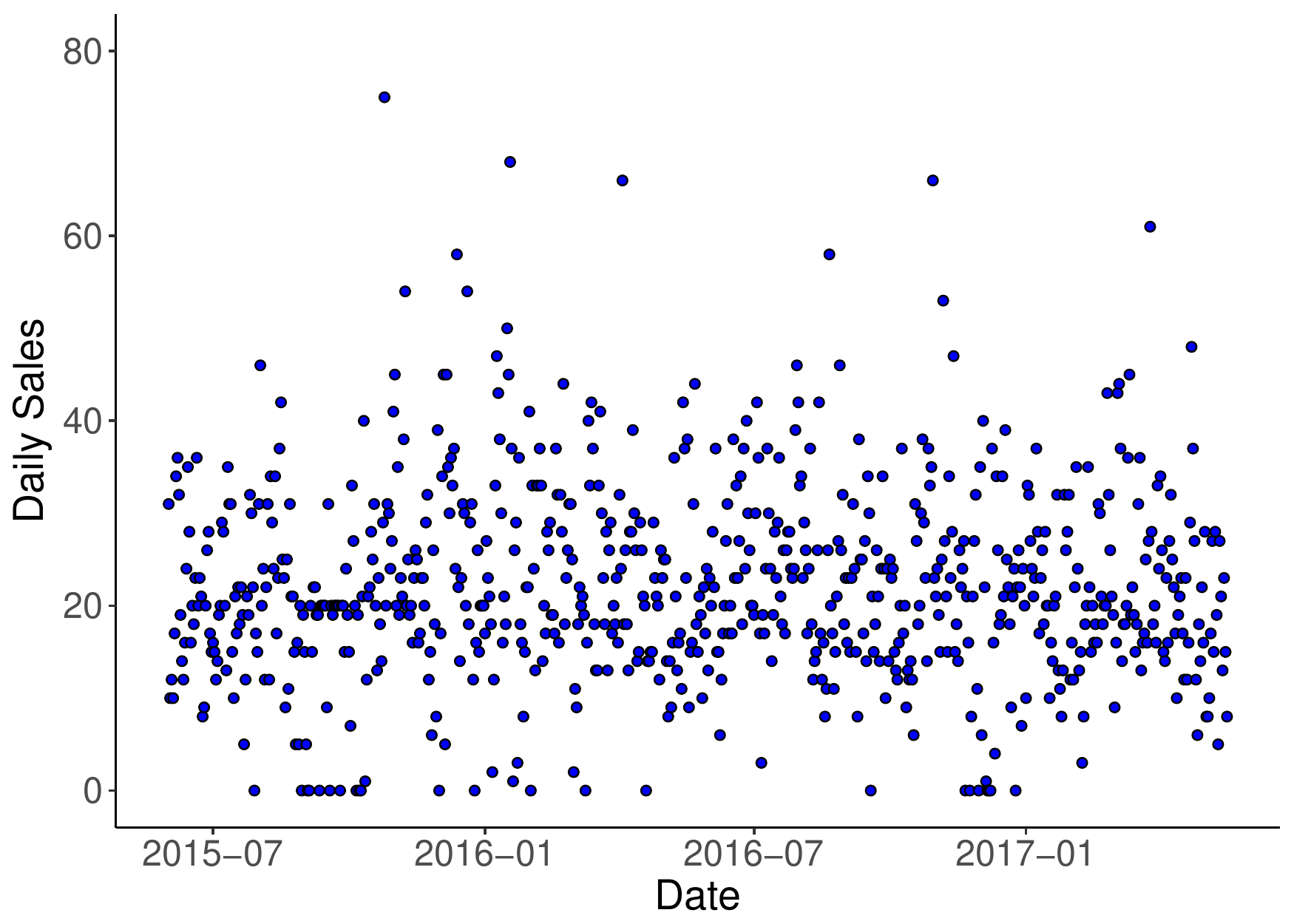} & 
		\includegraphics[width=.5\textwidth]{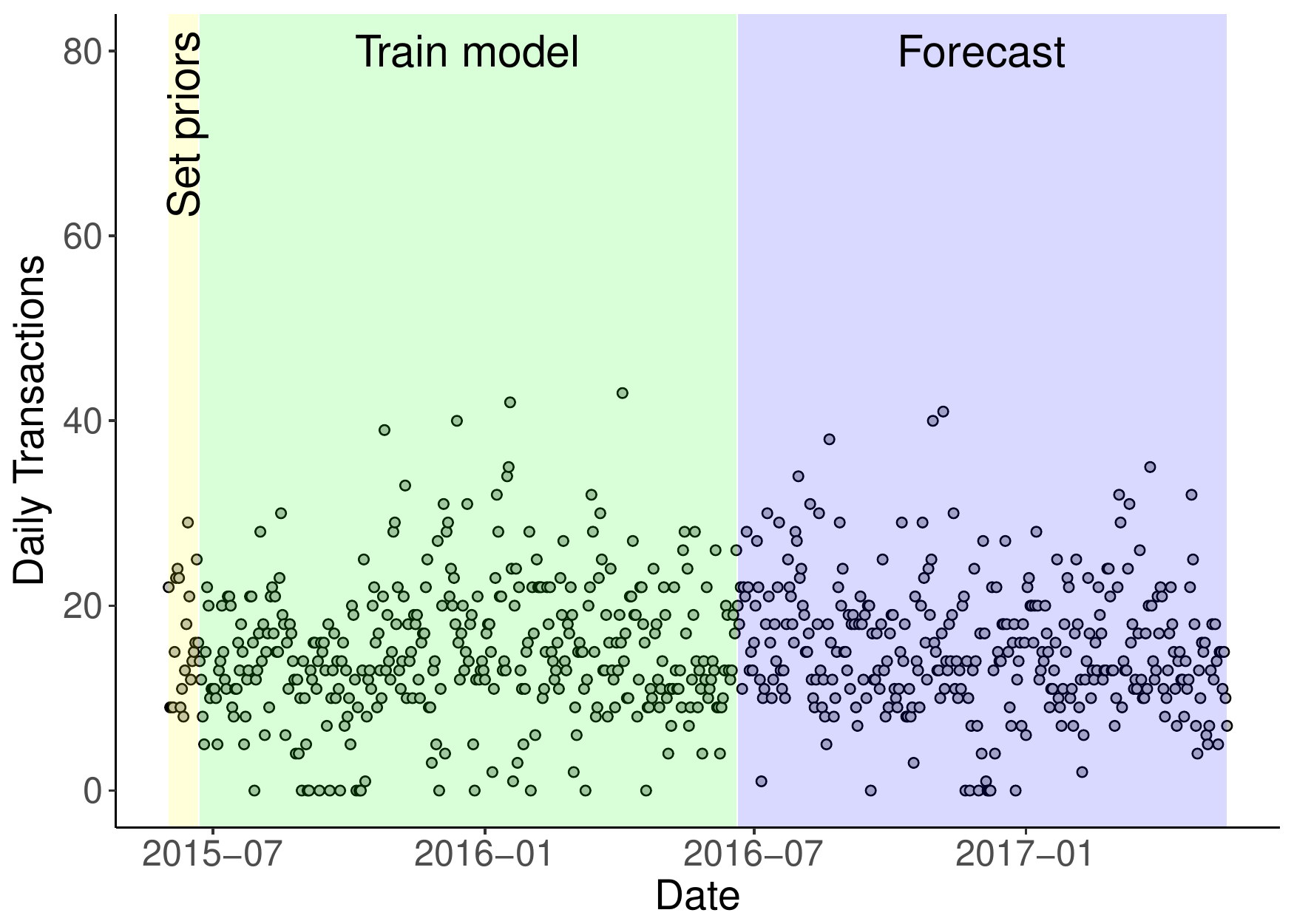} \\
		\multicolumn{2}{c}{(i) Item A}\\
		\includegraphics[width=.5\textwidth]{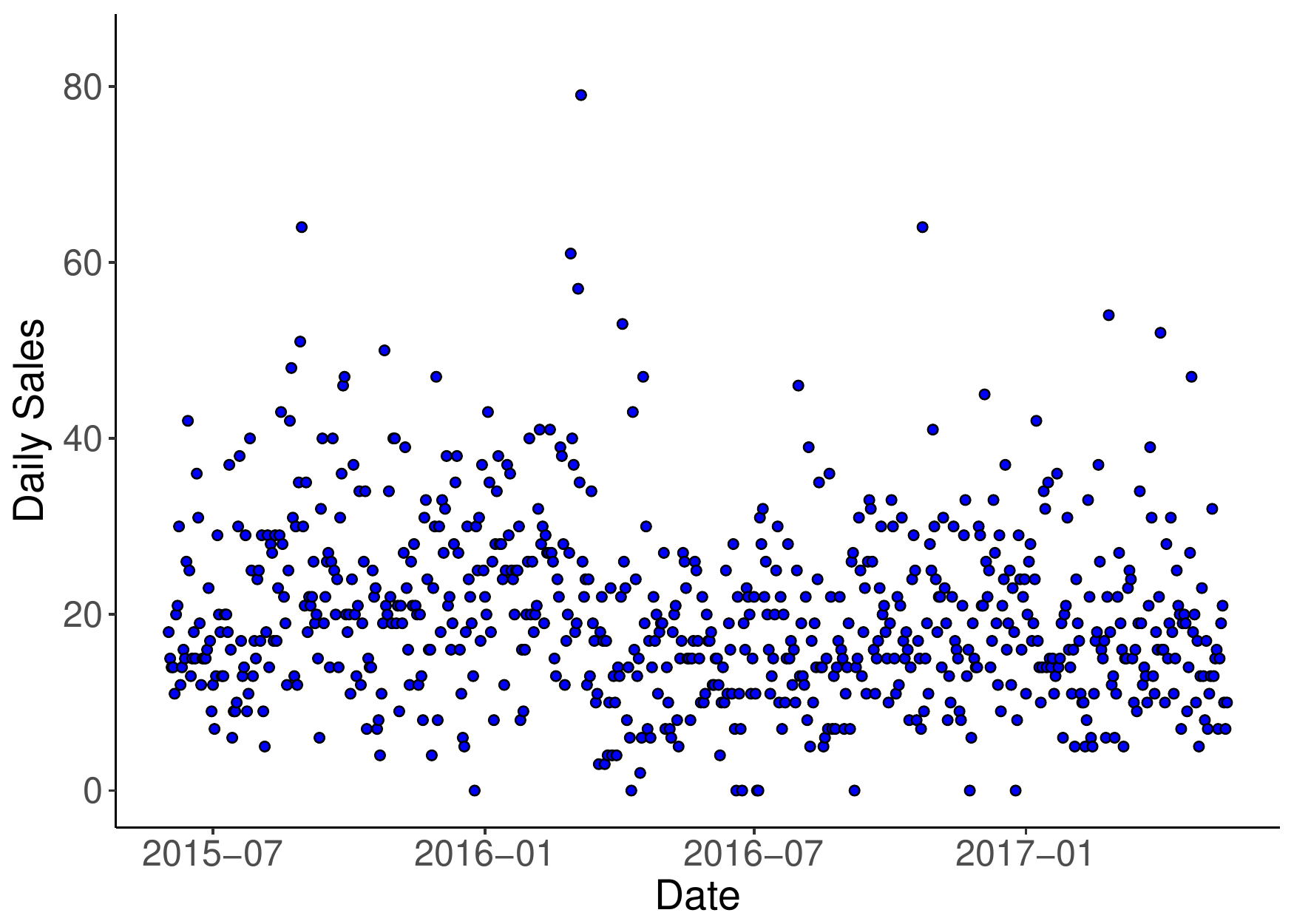} & 
		\includegraphics[width=.5\textwidth]{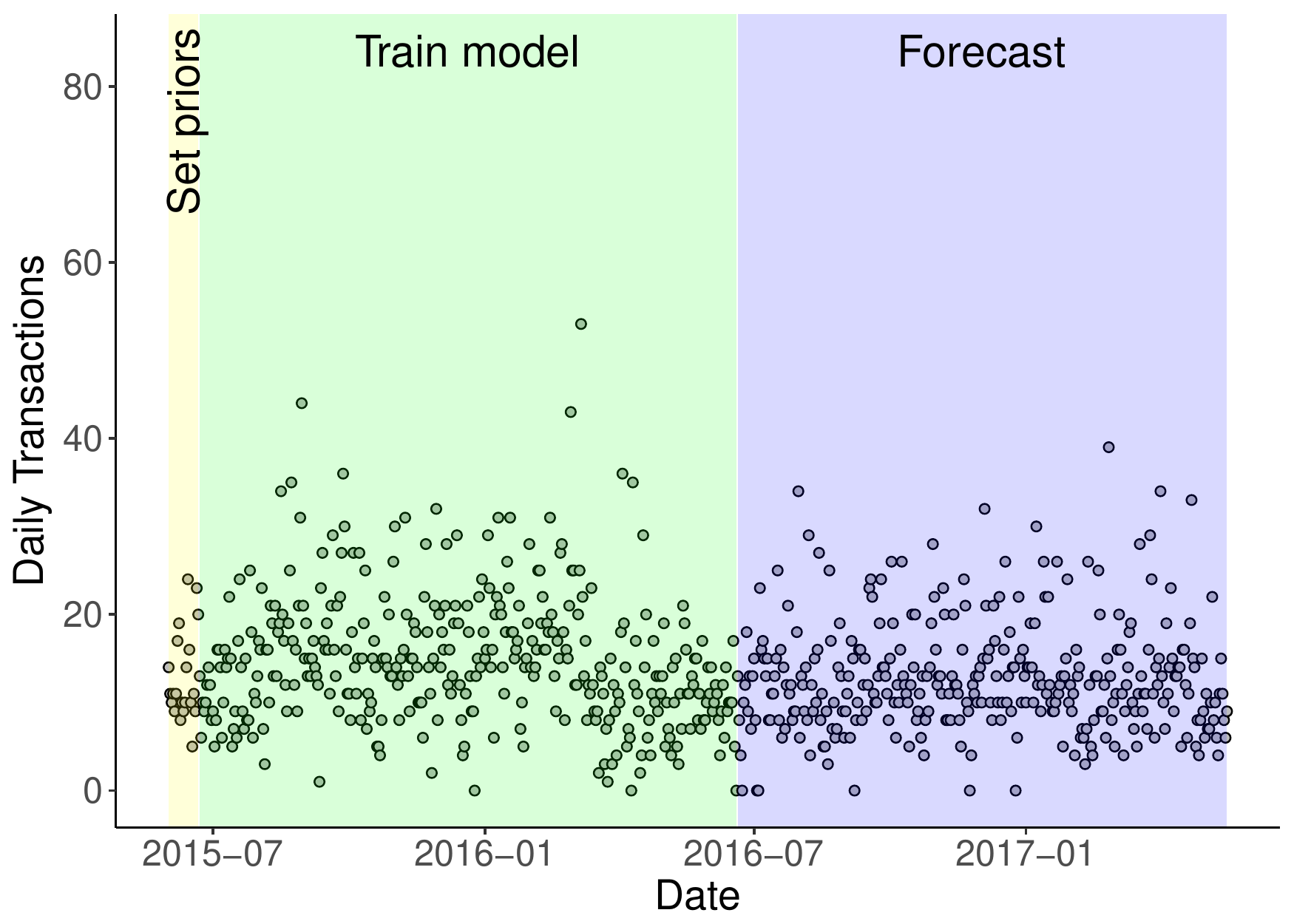}\\
		\multicolumn{2}{c}{(ii) Item B}\\
		\includegraphics[width=.5\textwidth]{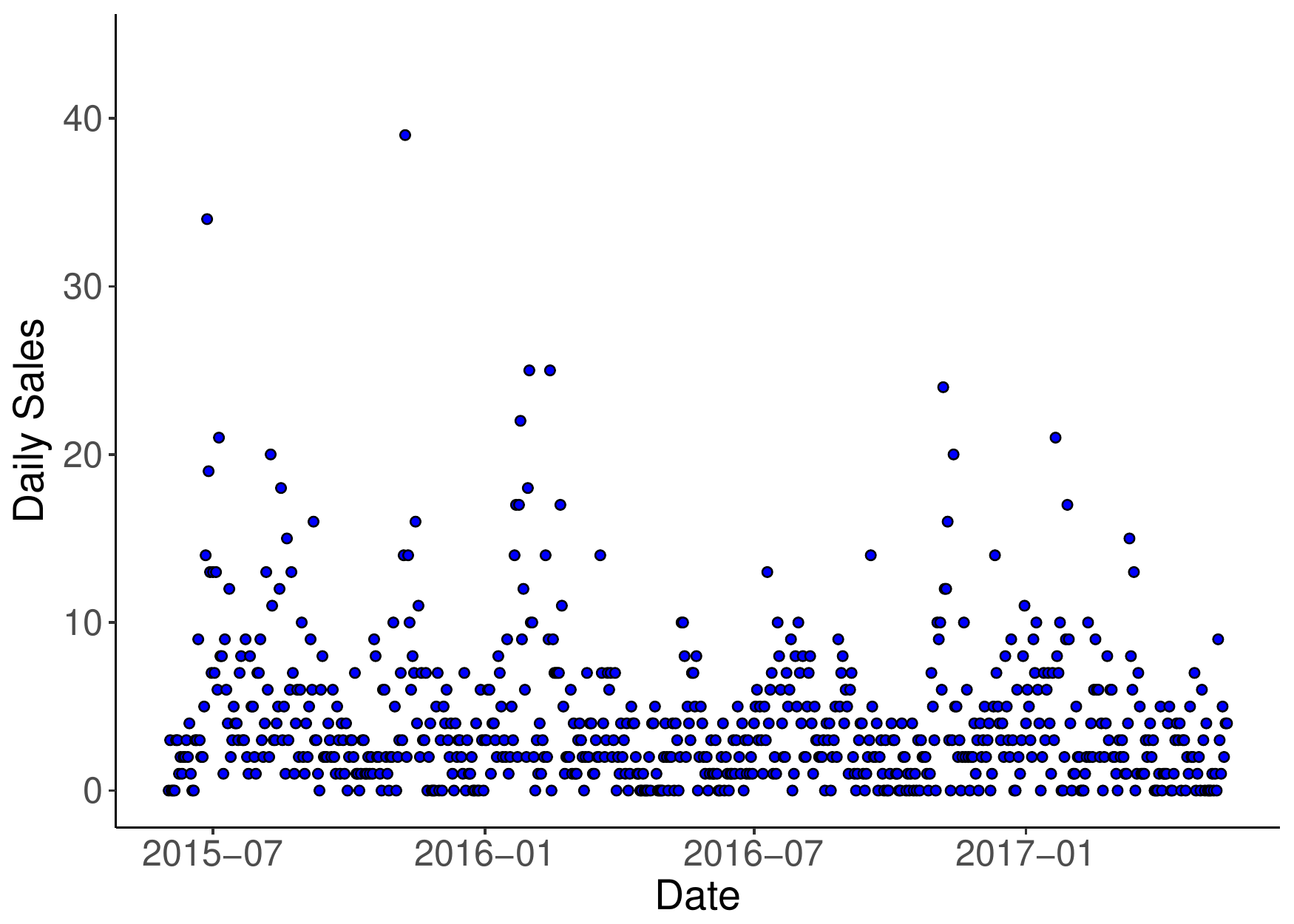} & \includegraphics[width=.5\textwidth]{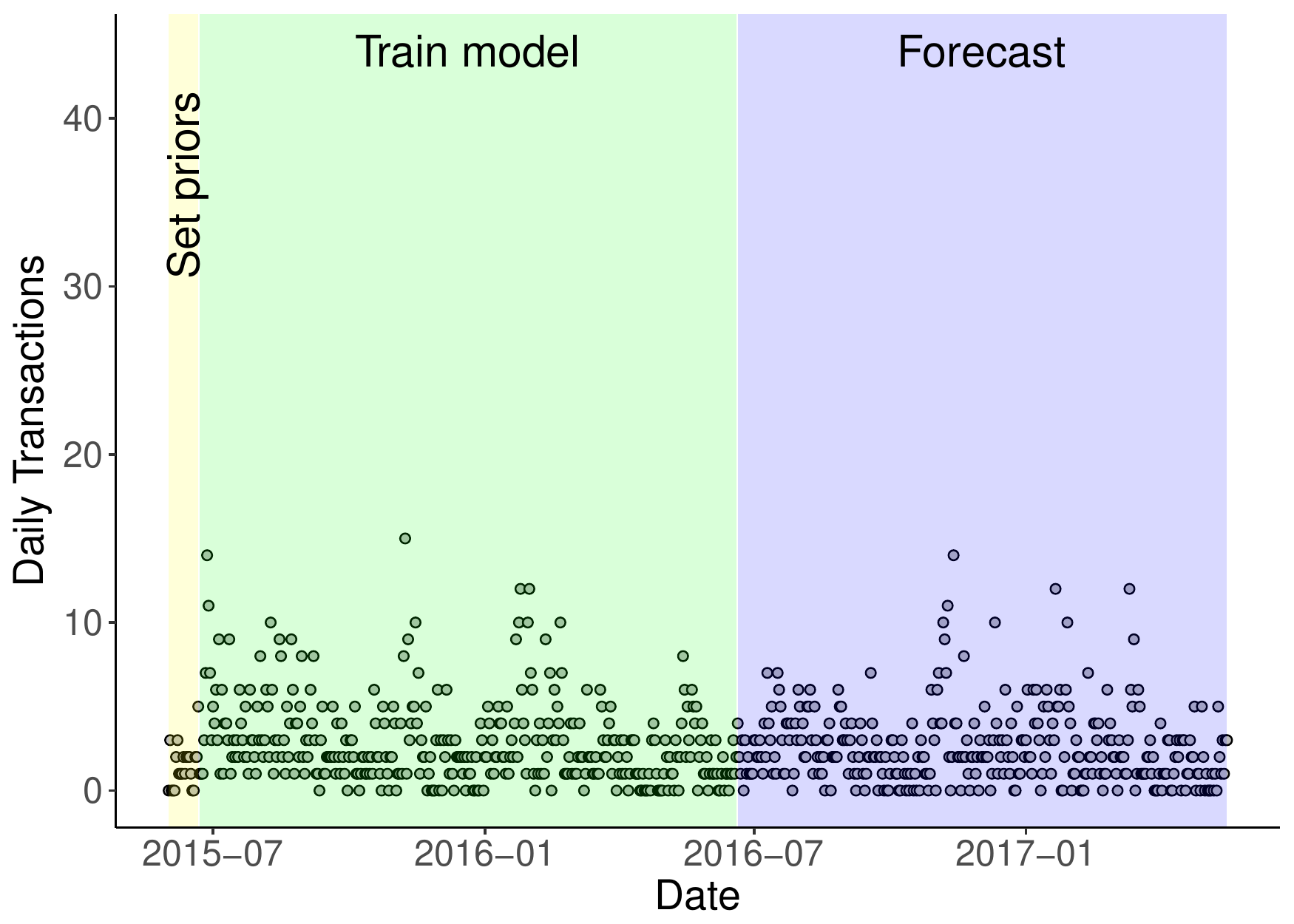}\\ 
		\multicolumn{2}{c}{(iii) Item C}\\
	\end{tabular}
	\caption{Daily sales and transactions of three spaghetti items (A-C) sold in one store from June 1st 2015 to July 1st 2017.}\label{fig:items}
\end{figure}
 
We explore forecasting of three spaghetti items to illustrate the potential improvements offered by decomposing heterogeneity into transactions and sales-per-transaction. These items represent a range of transactions-sales patterns and typify the features of data across many items. Table~\ref{tab:data} reports summaries of the daily transactions and sales-per-transaction for item A,B and C. Figure~\ref{fig:items} displays the daily transactions and sales for each item to illustrate the diminished diversity of item-level daily transactions in comparison to daily sales. Within this chosen category and store, items A and B are moderate to high selling items, and item C is a relatively low-selling item. Each item's daily sales and transactions share similar features such as the overall level and trends over time, and the evident day-of-week effect. Both series also share the feature of somewhat rare extreme values, although the diminished variability of the transaction data is evident.

\begin{table}[ht!]
	\centering
	\begin{tabular}{l|ccc|ccc}
		\hline
		& \multicolumn{3}{c|}{Daily transactions} & \multicolumn{3}{c}{Sales-per-transaction} \\
		Item & Mean & Median & Variance & Mean & Median & $\% < 5$ \\
		\hline
		A & 22.84 & 21 & 100.52 & 1.46 & 1 & 98.9 \\
		B & 19.75 & 18 & 101.15 & 1.44 & 1 & 99.0 \\
		C & 4.66 & 3 & 18.70 & 1.53 & 1 & 98.4 \\
		\hline
	\end{tabular}
\caption{Some summaries of daily transactions and sales-per-transaction data for 3 spaghetti items.}\label{tab:data}
\end{table}

\subsection{Model Specification \label{sec:modelsandpriors} }

\subsubsection{Transactions DCMM Specification} 
As described in Section~\ref{sec:DCMM}, the DBCM framework utilizes a DCMM to forecast daily transactions. In this analysis, we consider two DCMMs for forecasting transactions: independent DCMMs with item-specific weekly seasonal effects, and a multi-scale DCMM that shares information on the weekly seasonal effect across all spaghetti items.

The same form of DCMM is specified in the independent DBCM framework and the benchmark DCMMs on daily sales. In these independent DCMMs, each Bernoulli and conditionally Poisson component includes a local level, a full Fourier form seasonal component with period $7$, and a regression component with log price and a binary indicator of promotions as predictors. 
Each binary and conditionally Poisson DGLM can be defined through regression vectors and state evolution matrices of the form
$$\F'_t = \begin{pmatrix} 1, \, \log(\textrm{price}_t),\, \textrm{promo}_t, \, 1,0, \, 1,0, \, 1,0 \end{pmatrix}
\qquad\textrm{and}\qquad
\G_t = \text{blockdiag}[1,\, 1,\, 1, \, \H_1,\, \H_2,\, \H_3]
$$ 
with
$$\H_{j} = \begin{pmatrix} \phantom{-}\cos(2\pi j/7)  & \sin (2\pi j/7) \\
-\sin (2\pi j/7) & \cos(2 \pi j/7)\end{pmatrix},\quad j=\seq13.$$
where $\textrm{price}_t$ is the item-specific price on day $t$, and $\textrm{promo}_t$ is equal to 1 if the item is on promotion on day $t$, and 0 if not. Through the standard use of discount factors, each component is dynamic, allowing for time variation in the level, weekly seasonality, and price and promotion effects. Based on previous analyses of item-level sales and transactions, we set fixed discount factors of 0.99 (Poisson) and 0.999 (Bernoulli) on each component. 

The multi-scale DBCM includes item-level models $\cM_i$ with $\f'_{i,t} = \begin{pmatrix} 1, \, \log(\textrm{price}_{i,t}), \, \textrm{promo}_{i,t} \end{pmatrix}$, and a scalar factor $\phi_t$ representing the current day-of-week effect. In this multi-scale analysis, $\cM_0$ is a dynamic linear model (DLM) on the aggregate log daily transactions of all spaghetti items in the chosen store. This aggregate DLM includes a local linear trend, the scaled log average spaghetti price as a predictor, and full Fourier form seasonal components of periods $7$ and $365$ representing the weekly and yearly seasonal effects. We allow for dynamic level, trend, regression effects, and seasonality with discount factors of $\delta = 0.995$ for the trend and regression components, $\delta = 0.999$ for each of the seasonal components, and $\beta = 0.999$ for the residual stochastic variance process. Predictive performance in all sales/transactions DCMMs is evaluated across a range of random effects discount factors, $\rho \in (.2, .4, .6, .8, 1)$. 

The shading in Figure~\ref{fig:items} indicates analysis set-up. For each DCMM and the aggregate DLM, initial priors using three weeks of training data (yellow shading). For the aggregate log-normal DLM and the conditionally Poisson DGLMs, we define approximate prior moments for the state vectors based on the posterior moments in a standard reference analysis of a Bayesian linear model of the log daily sales/transactions. For the binary DGLMs, we estimate the prior mean of the level to be $\log(p/(1-p))$, where $p$ is the observed proportion of the first 21 days with at least one transaction. All other prior means in the binary DGLM are set to zero, with the prior covariance matrix as the identity. The green shaded region in Figure~\ref{fig:items} denotes the one year period beginning on day 22 (denoted $t=1$) in which our models are trained. After this one year period, in the blue shaded region, forecasting $\seq{1}{14}$-days ahead is performed on each of the 332 days. 

\subsubsection{Binary Cascade Model Specification} Based on an exploratory analysis of typical sales-per-transaction, we set $d = 4$ for all items in this analysis. As seen in Table~\ref{tab:data}, around 99\% of all transactions of the chosen items include four or fewer unit sales. The form of the binomial logistic DGLMs is the same across items and for all $r = \seq{1}{d}$. Each conditional model of $n_{r,t}$ includes a dynamic local level, and a static regression component with a binary indicator of promotion as a predictor. Each binomial DGLM allows for slow time variation in the level through a discount factor of $\delta = 0.999$. In previous analyses, we found a static promotional effect, with $\delta = 1$, to be sufficient. For each binomial logistic DGLM, we specify, 
$$\F_{r,t}^{0} = \begin{pmatrix} 1, \, \textrm{promo}_t\end{pmatrix}'
\qquad\textrm{and}\qquad
\G_{r,t} = \textbf{I}
$$ 
where the $\textrm{promo}_t$ is an item-specific indicator of a promotion at time $t$. Again, we use three weeks of training data to specify the prior mean of the level. In a logistic model of $\pi_{r,t}$, we set the prior mean of the level to be $\log(p/(1-p))$ where $p$ is the proportion of transactions with exactly $r$ unit sales out of all transactions with at least $r$ unit sales. We set the prior mean of the promotion coefficient to be zero, and the prior covariance matrix for the state vector to $(.1) \mathbf{I}$. 

\subsubsection{Excess Distribution\label{sec:modelsandpriorsexcess}} 
We consider two perspectives: leaving the excess distribution completely unspecified, or bootstrapping from the empirical excess distribution. In this context of daily sales forecasting, unpredictable and relatively rare situations may arise where, for example, a consumer purchases dozens or hundreds of units in a single bulk order. Due to lack of relevant data and predictors that would make modeling these rare outcomes possible, it is often preferable to leave the tail of the sale-per-transaction distribution unspecified. However, without constraints or assumptions on the excess distribution, we are limited in the conclusions we can make about the predictive distribution. At time $t-1$,   the $1$-step forecast density of $y_t$ is
$$p(y_{t} \mid \cD_{t-1},\cI_{t-1}) = q_t f(y_{t}) + (1-q_t) p_d(y_{t})$$
where: (i)  $q_t = Pr(n_{d,t} > 0)$ is the probability that  $n_{d,t}>0$, i.e., that {\em some} of the transactions have more than $d$ units; (ii) $f(y_t)$ is the p.d.f. of the sales distribution given that  $n_{d,t}>0$;  and $p_d(y_t)$ is the p.d.f. of the (specified) distribution given that $n_{d,t} = 0. $    The forecast p.d.f.s for multi-steps ahead have similar forms.  If $f(\cdot)$ is unspecified, we cannot exactly identify the mean or quantiles of the distribution. It is possible to identify lower/upper bounds for any quantile of the forecast distribution, including the median, but without additional assumptions about $f$, bounds on the mean of the forecast distribution are not available.

The second perspective is to utilize the empirical distribution of excess sales over a past period of time. Simulating excess sales-per-transaction from the empirical excess distribution results in access to the entire predictive distribution through Monte Carlo samples. With this approach, we can report any quantity of interest from the forecast distribution. Since forecasters are often interested in the accuracy of many different error metrics (and the corresponding optimal point forecasts), we present the results of the DBCM models using the empirical excess distribution. A potential downside of this approach is that the only possible values of sales-per-transaction are those that have previously been observed; that excesses are very rare ameliorates this concern. Other specifications that may be of utility are noted in the concluding section. 

\subsection{Examples and Evaluations  \label{sec:examples} }
 
\subsubsection{Joint Forecast Trajectories and Probabilistic Evaluation}
 
\begin{figure}[hp]
 	\centering
 	\begin{tabular}{cc}
 		\textbf{Mar 20 2017} & \textbf{Apr 25 2017}\\ 
 		\includegraphics[width=.5\textwidth]{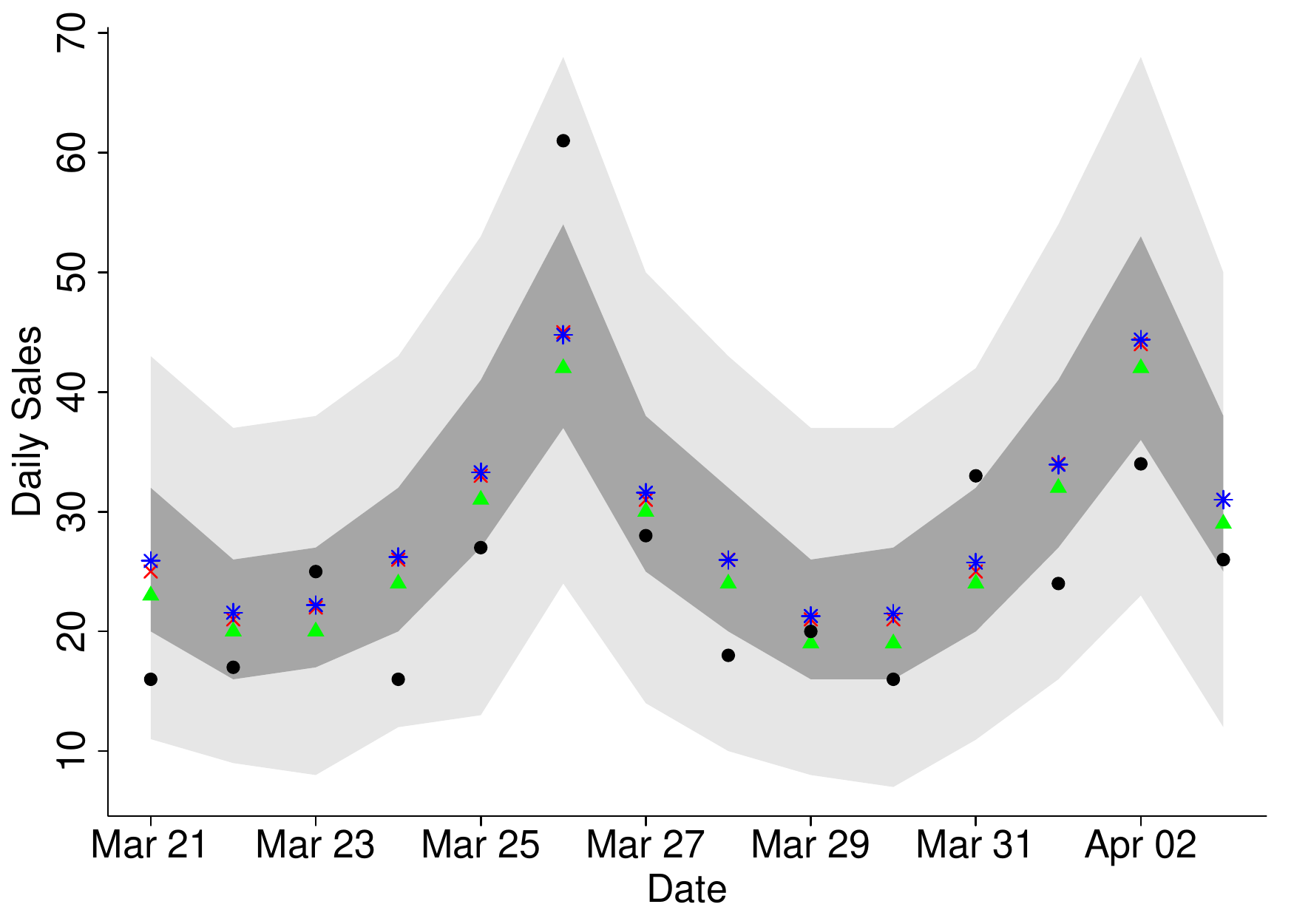} & 
 		\includegraphics[width=.5\textwidth]{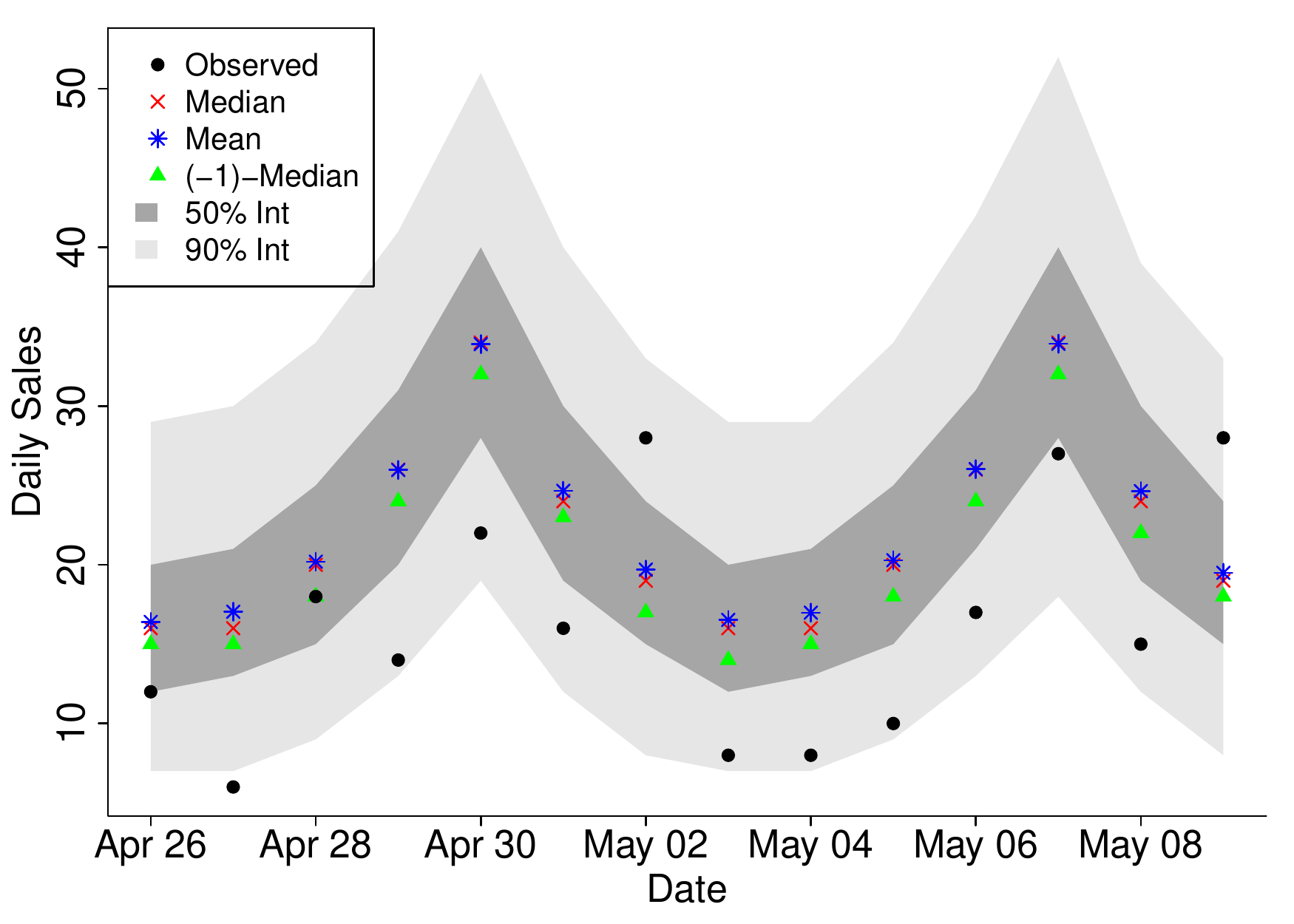} \\
 		\multicolumn{2}{c}{(i) Item A} \\
 		\includegraphics[width=.5\textwidth]{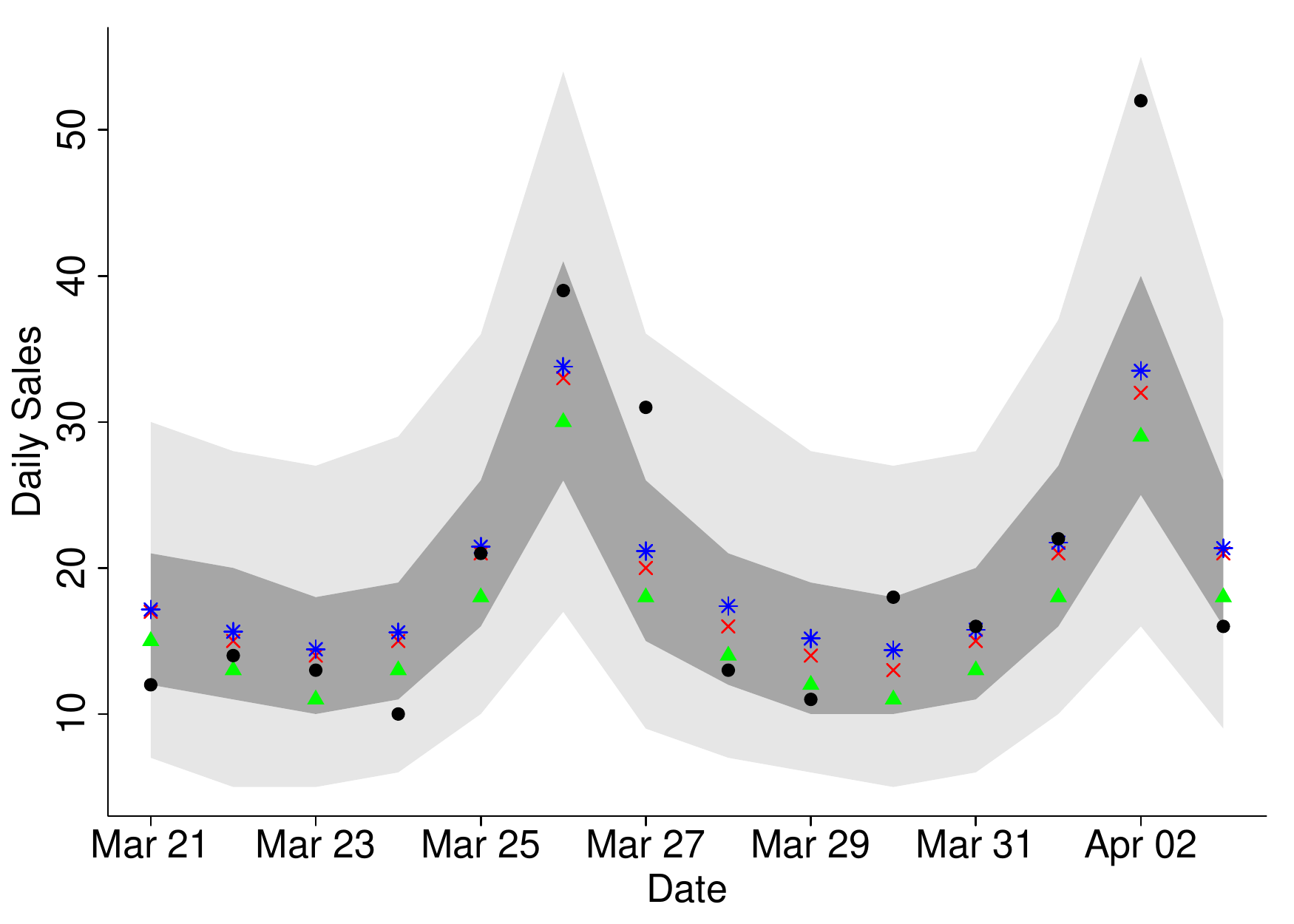} & 
 		\includegraphics[width=.5\textwidth]{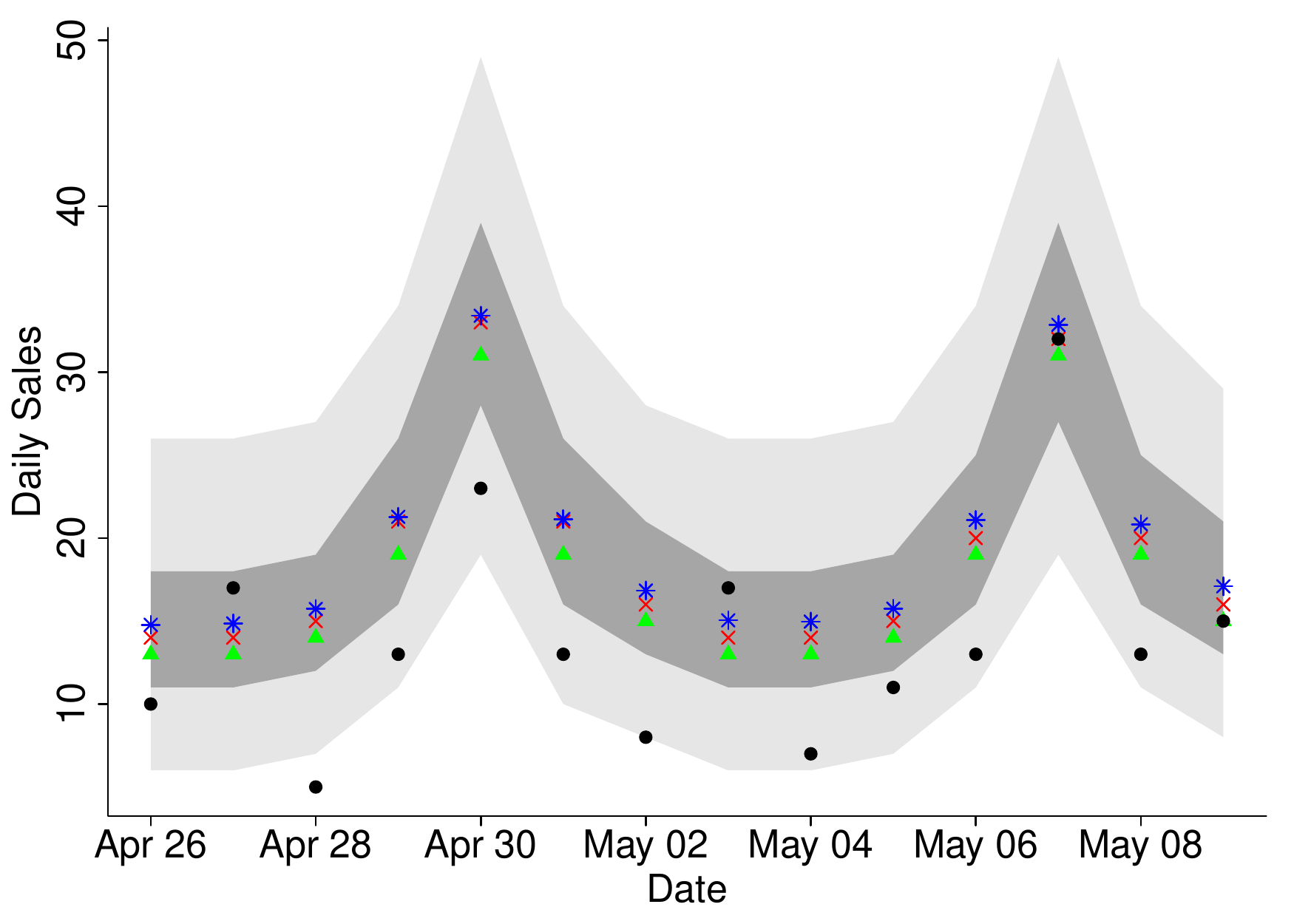} \\
 		\multicolumn{2}{c}{(ii) Item B} \\
 		\includegraphics[width=.5\textwidth]{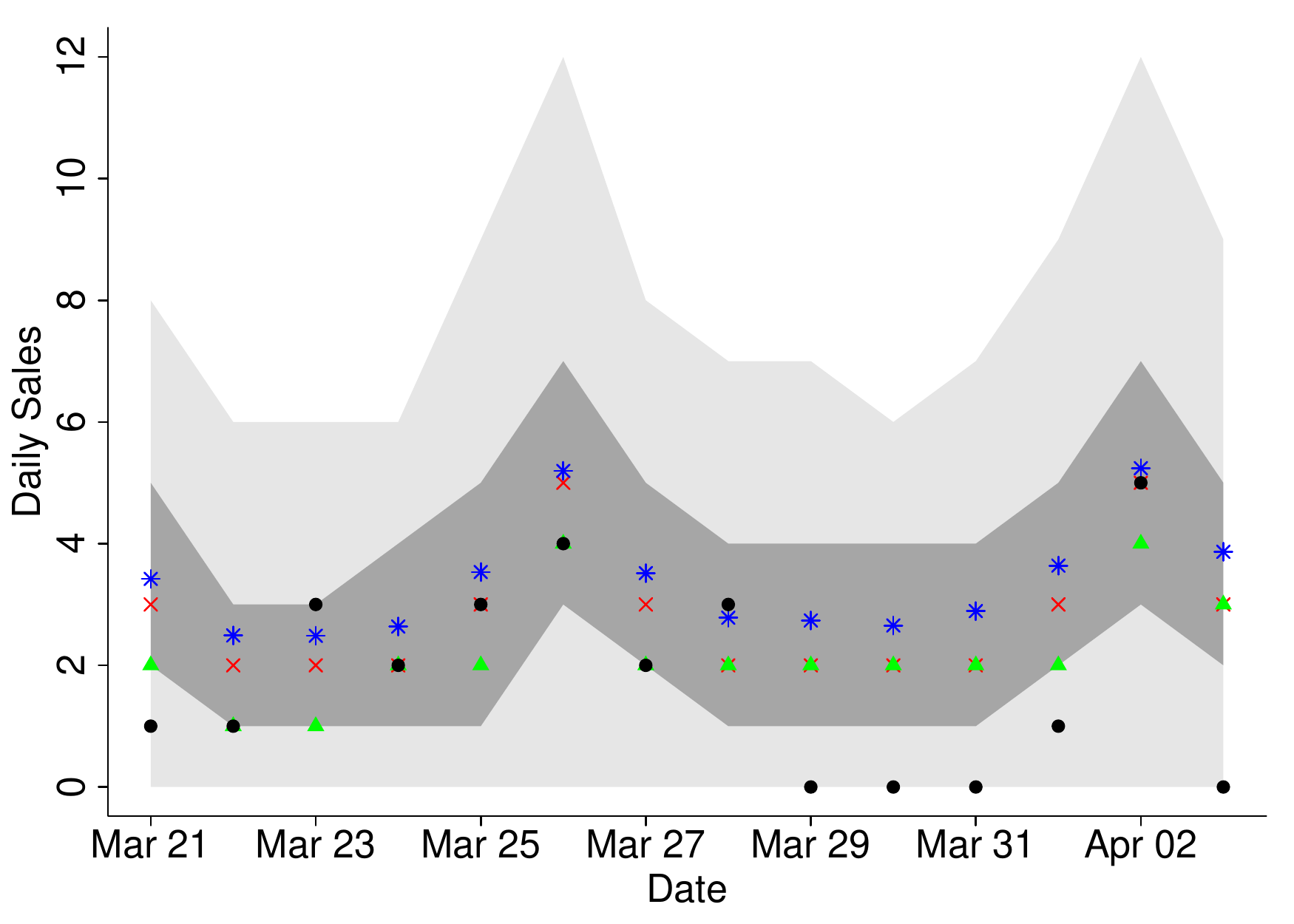} & 
 		\includegraphics[width=.5\textwidth]{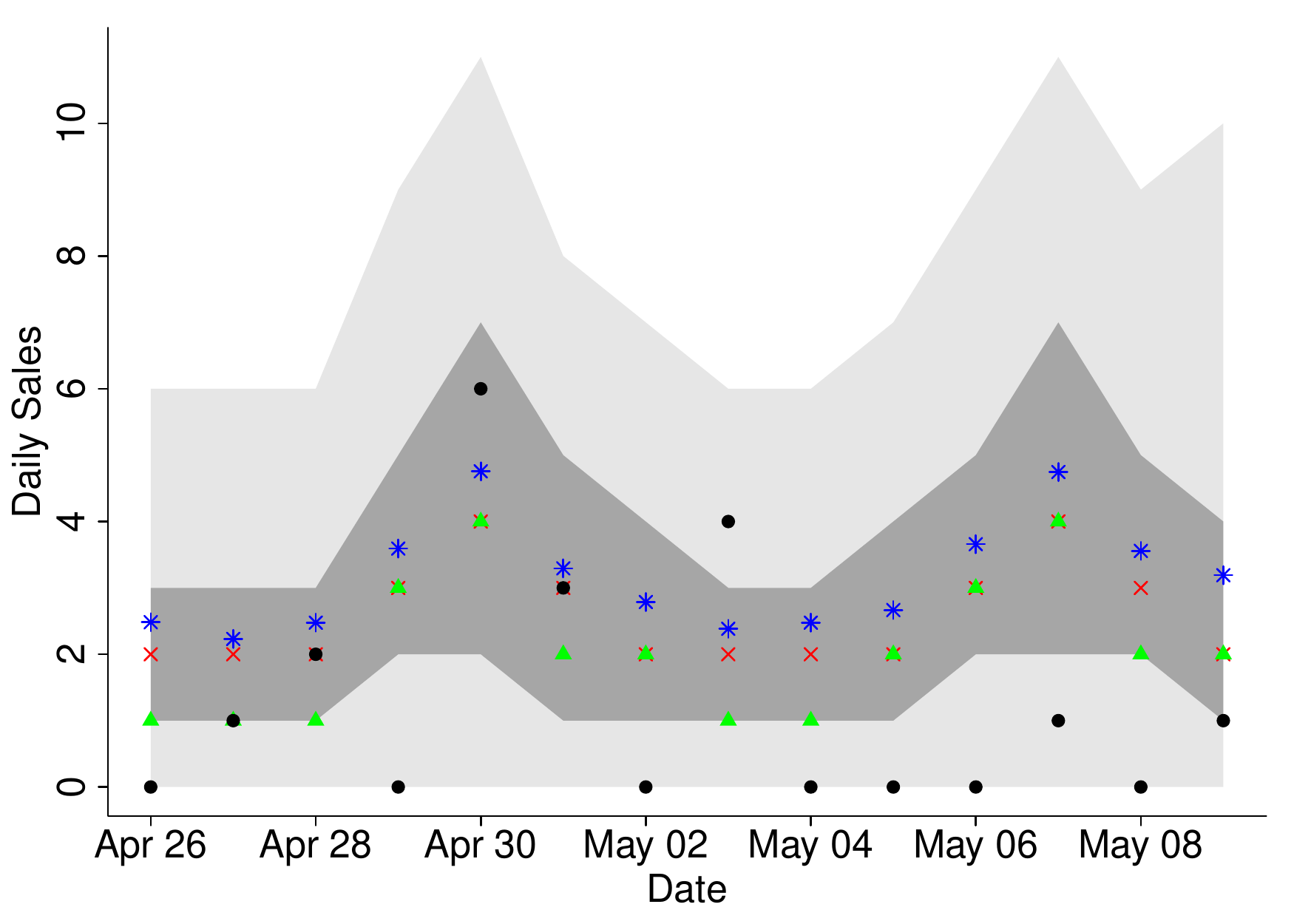}  \\ 
 		\multicolumn{2}{c}{(iii) Item C} \\
 	\end{tabular}
 \caption{1-14 day joint forecast trajectories on Mar 20th 2017 (left) and Apr 25 2017 (right). Observed daily sales shown as a circle, forecast median as an x, forecast mean as a diamond, and forecast $(-1)$-median as a triangle. Light and dark shading indicate the forecast 50 and 90\% credible intervals, respectively. }\label{fig:trajectories}
 \end{figure}

Example forecast trajectories from this analysis are shown in Figure~\ref{fig:trajectories}. These plots illustrate $\seq{1}{14}$-day ahead joint forecasts on two days, Mar 20th 2017 (left column) and Apr 25th 2017 (right column). For each item, these forecasts were generated from the multi-scale binary cascade model, and the excess sales was drawn from empirical excess distribution. The displayed forecasts from the DBCM model are based on transaction forecasts from a DCMM with a random effects discount factor of $\rho = 1$. 
These plots provide insight into the spread of the forecast distribution ($50, 90\%$ credible intervals in gray shading), as well as the location of common point forecasts (mean, median, and $(-1)$-median). Observed daily sales  are shown as black circles. 

In general, forecasts made on Mar 20th were accurate in terms of location and spread. For item A, $7/14$ days are contained in the $50\%$ credible intervals, and $14/14$ in the $90\%$ intervals. For item B, the $50\%$ intervals contain $11/14$ days, and the $90\%$ intervals contain $14/14$ days. For item C, the $50\%$ intervals contain $8/14$ days, and the $90\%$ intervals contain $14/14$ days. On Apr 25th, the point forecasts are somewhat over-estimates,  while  $50\%$ intervals show some under-coverage. For items A, B, and C, the $50\%$ intervals contain only $2/14$, $4/14$, and $5/14$ days, respectively. However,  $90\%$ intervals for each item are more accurate, containing $13/14$, $13/14$, and $14/14$ observations, respectively. These trajectories simply provide   snapshots of  forecasts on two single days, to highlight the underlying forecasting process; coupled with this, we now evaluated aspects of longer-term forecasting performance.

    \begin{figure}[hp]
 	\centering
 	\begin{tabular}{cc}
 		\textbf{Coverage} & \textbf{PIT}\\ 
 		\includegraphics[width=.35\textwidth]{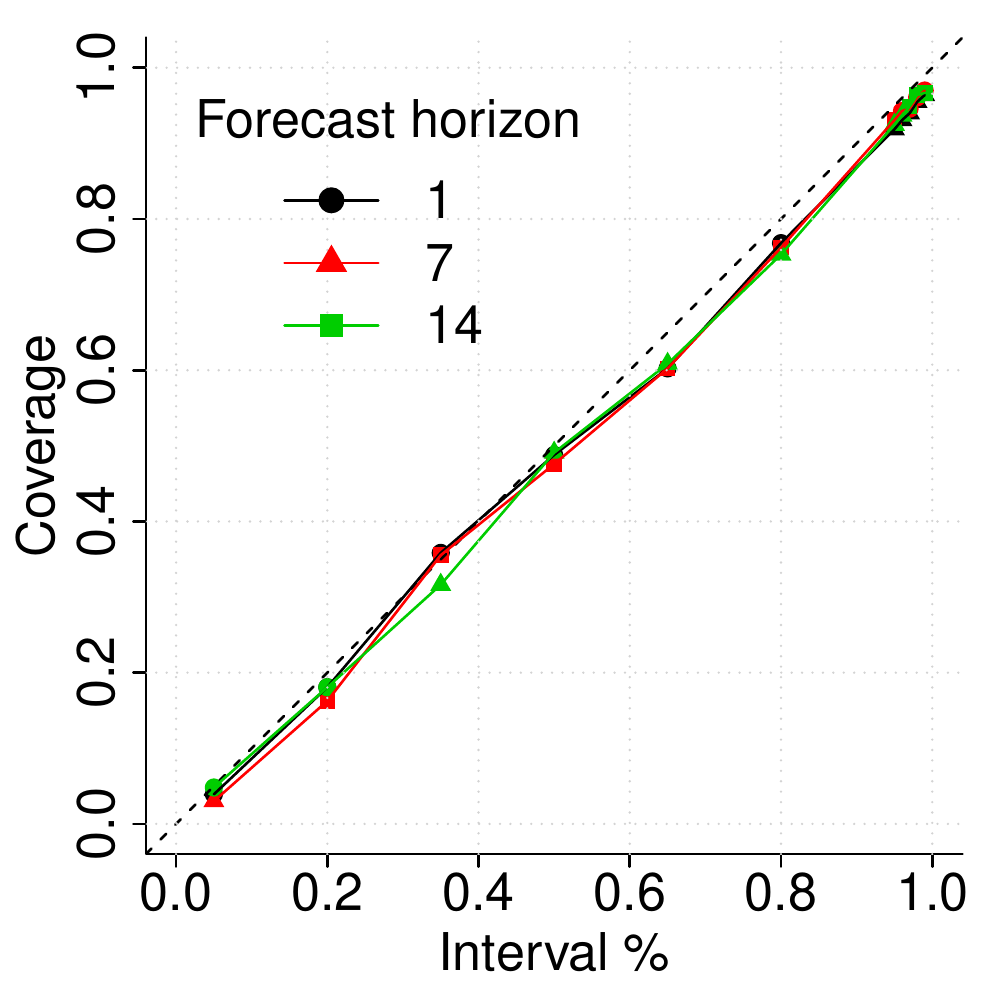} & 
 		\includegraphics[width=.35\textwidth]{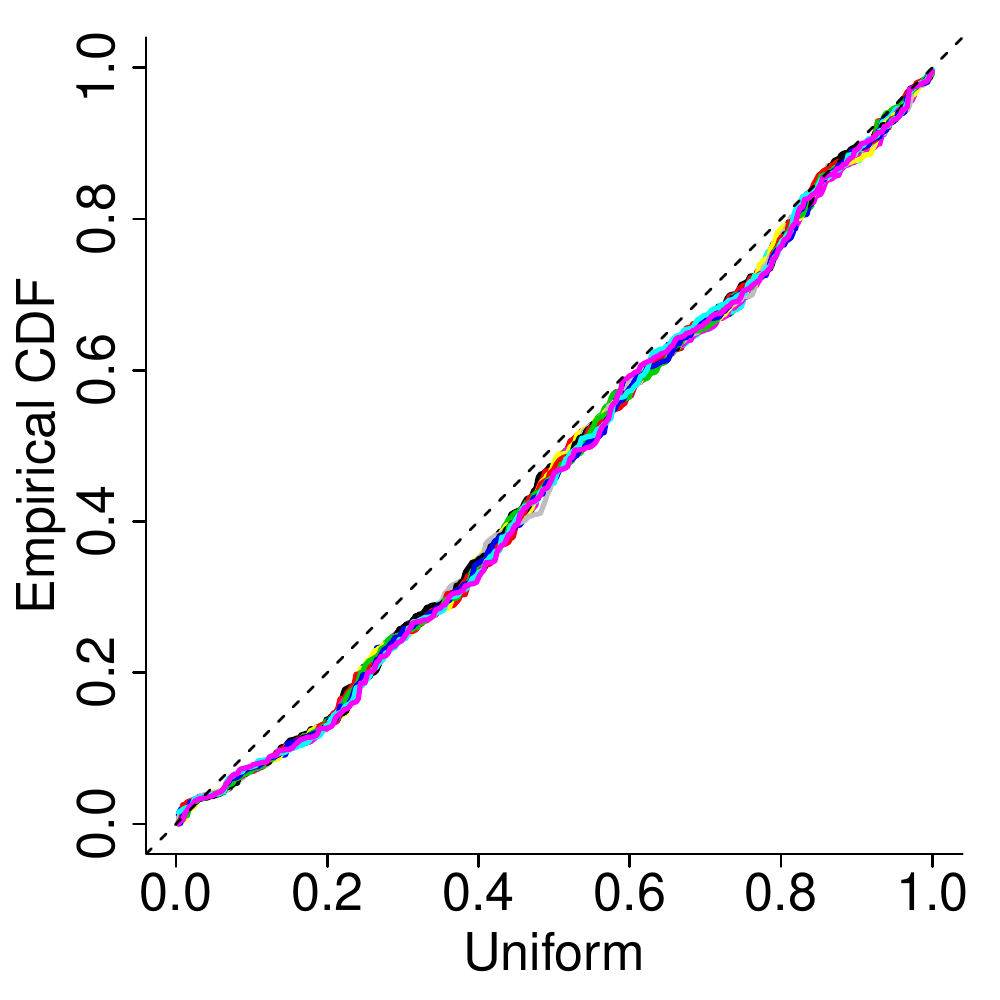} \\
 		\multicolumn{2}{c}{(i) Item A} \\
 		\includegraphics[width=.35\textwidth]{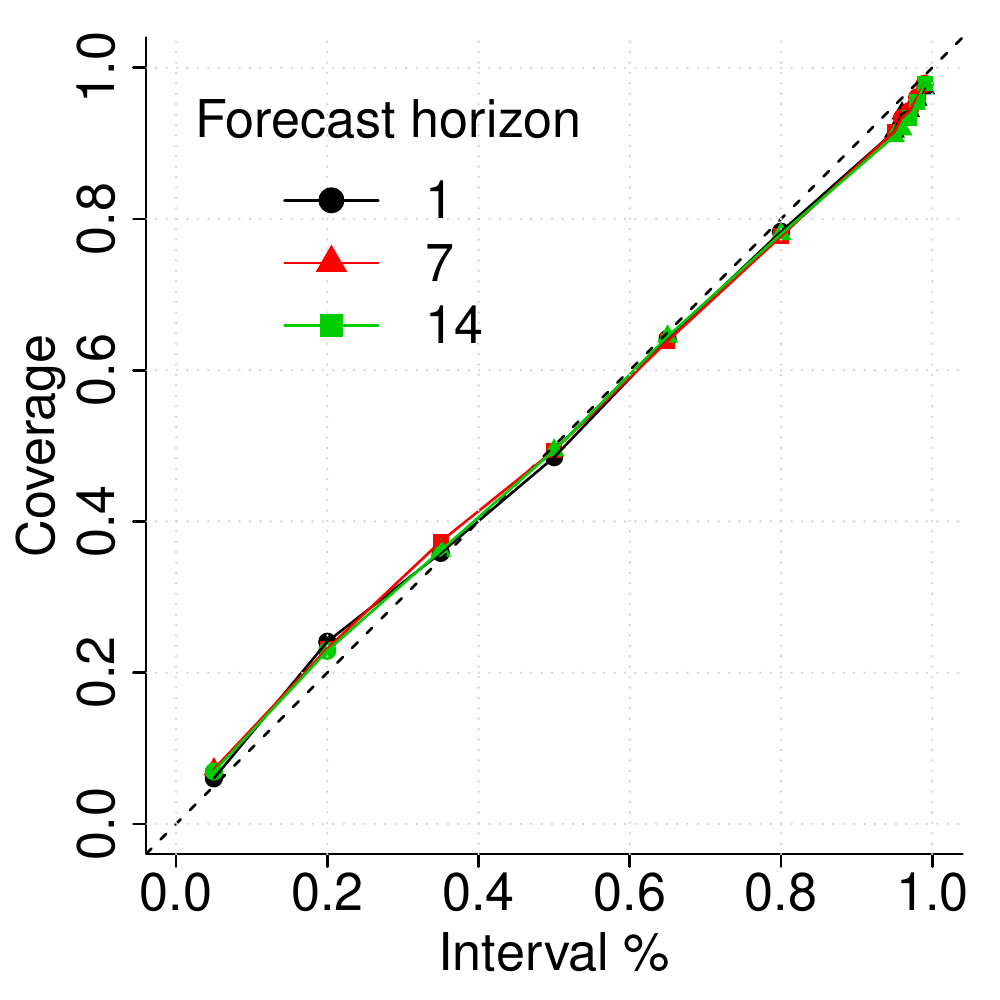} & 
 		\includegraphics[width=.35\textwidth]{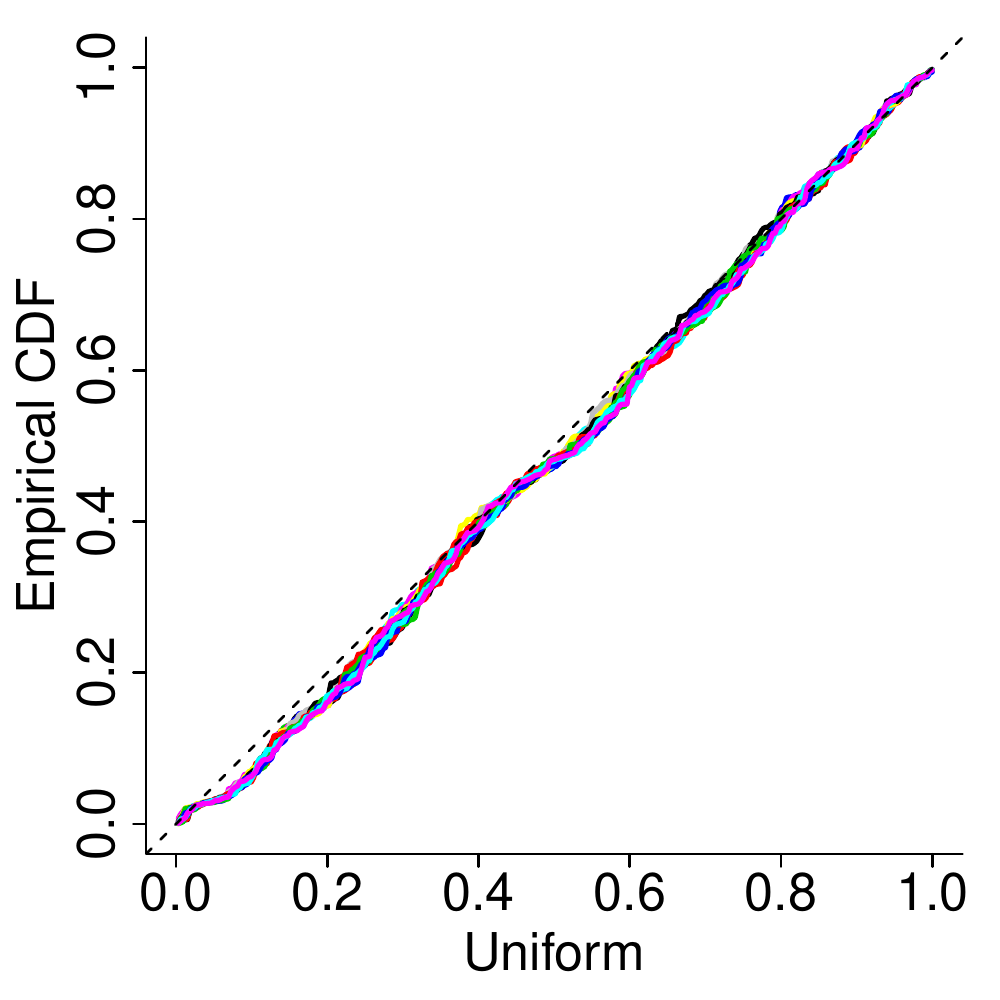} \\
 		\multicolumn{2}{c}{(ii) Item B} \\
 		\includegraphics[width=.35\textwidth]{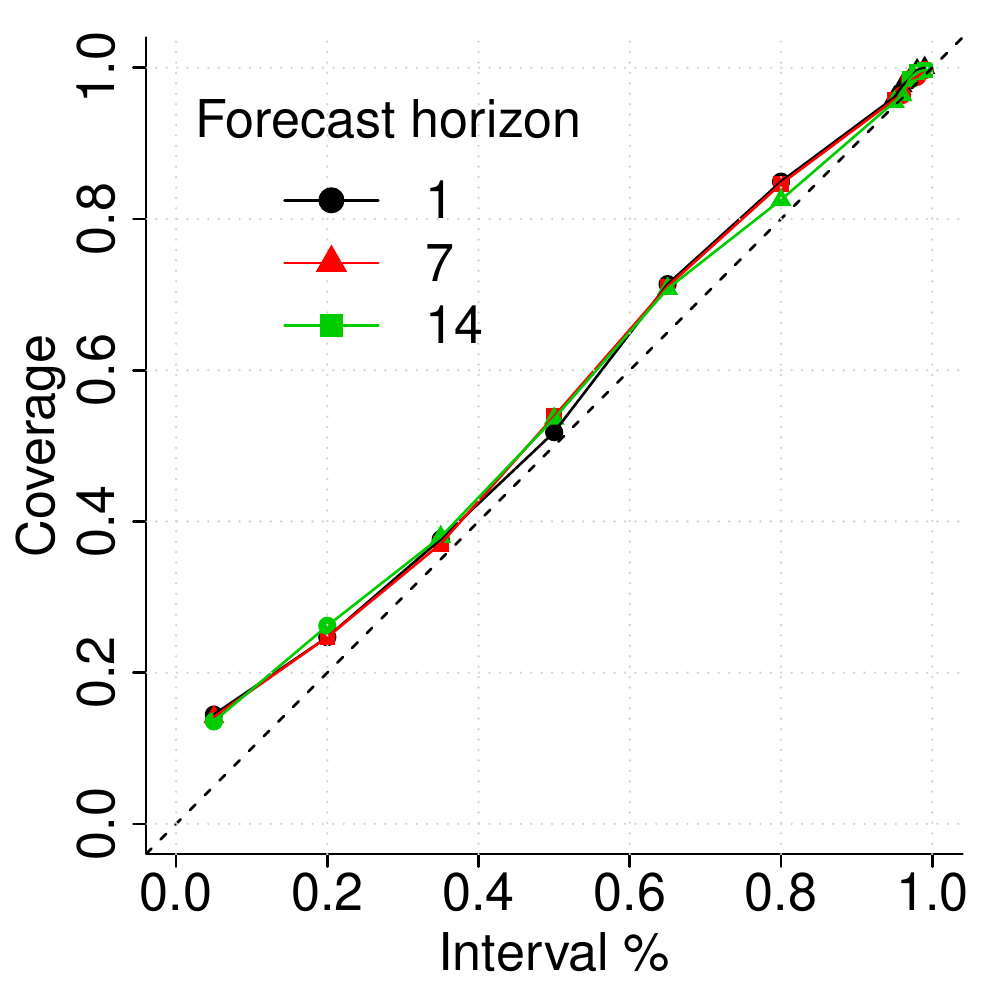} & 
 		\includegraphics[width=.35\textwidth]{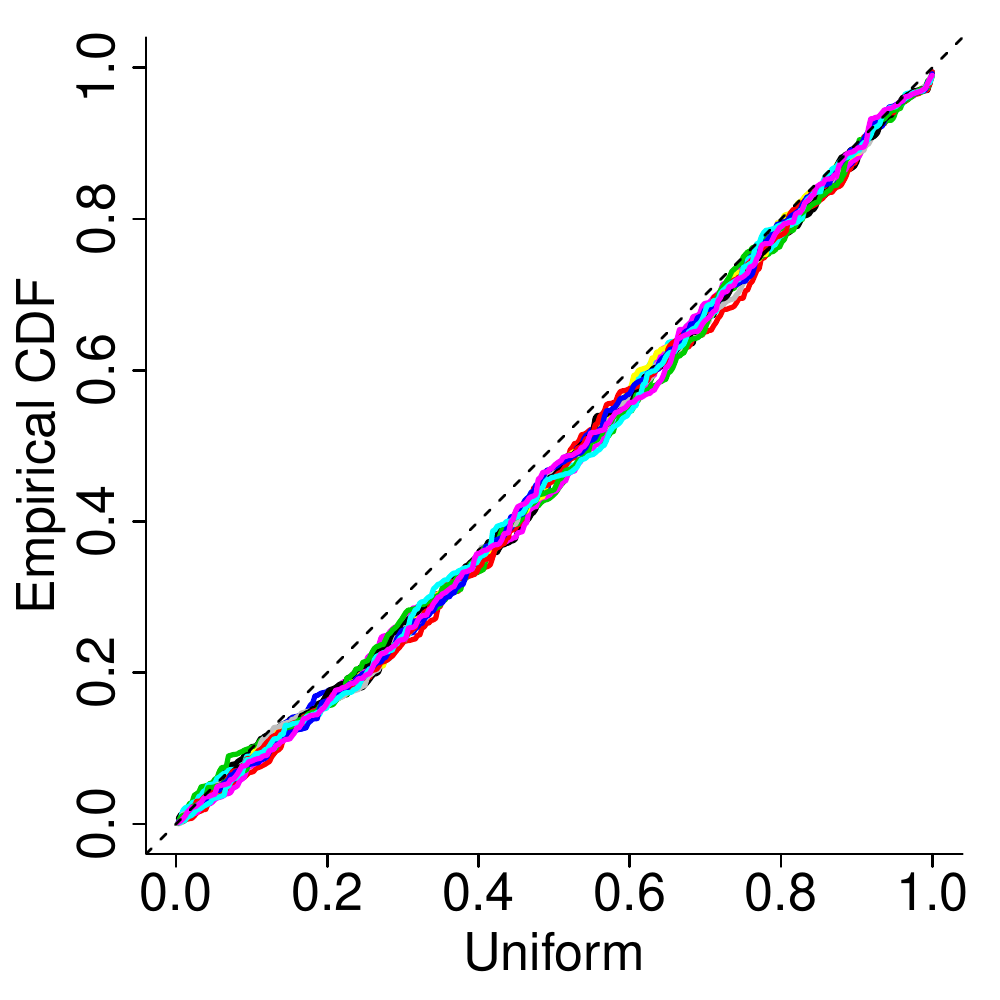}  \\ 
 		\multicolumn{2}{c}{(iii) Item C} \\
 	\end{tabular}
 	\caption{Empirical coverage plots (left) for $1, 7, 14$-step forecasts and randomized PIT plot (right) for 1-14 step forecasts of items A (top), B (middle), and C (bottom) using the multi-scale DBCM with empirical excess distribution and random effect discount factor of $\rho = 1$.}\label{fig:distr_eval}
 \end{figure}

  \begin{figure}[p]
  	\centering
 	\begin{tabular}{cc}
 		\textbf{MAD} & \textbf{MAPE}\\ 
 		\includegraphics[width=.5\textwidth]{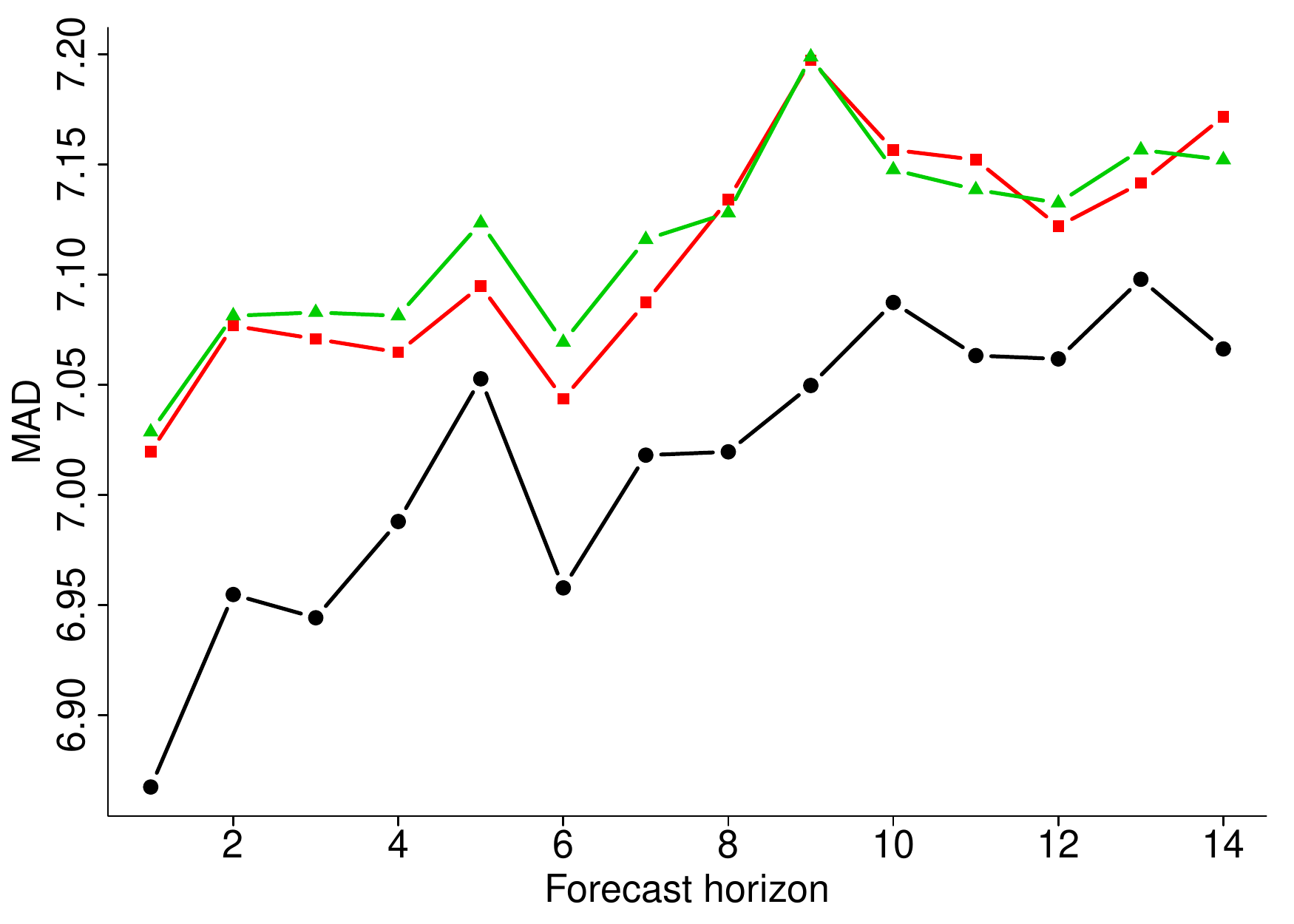} & 
 		\includegraphics[width=.5\textwidth]{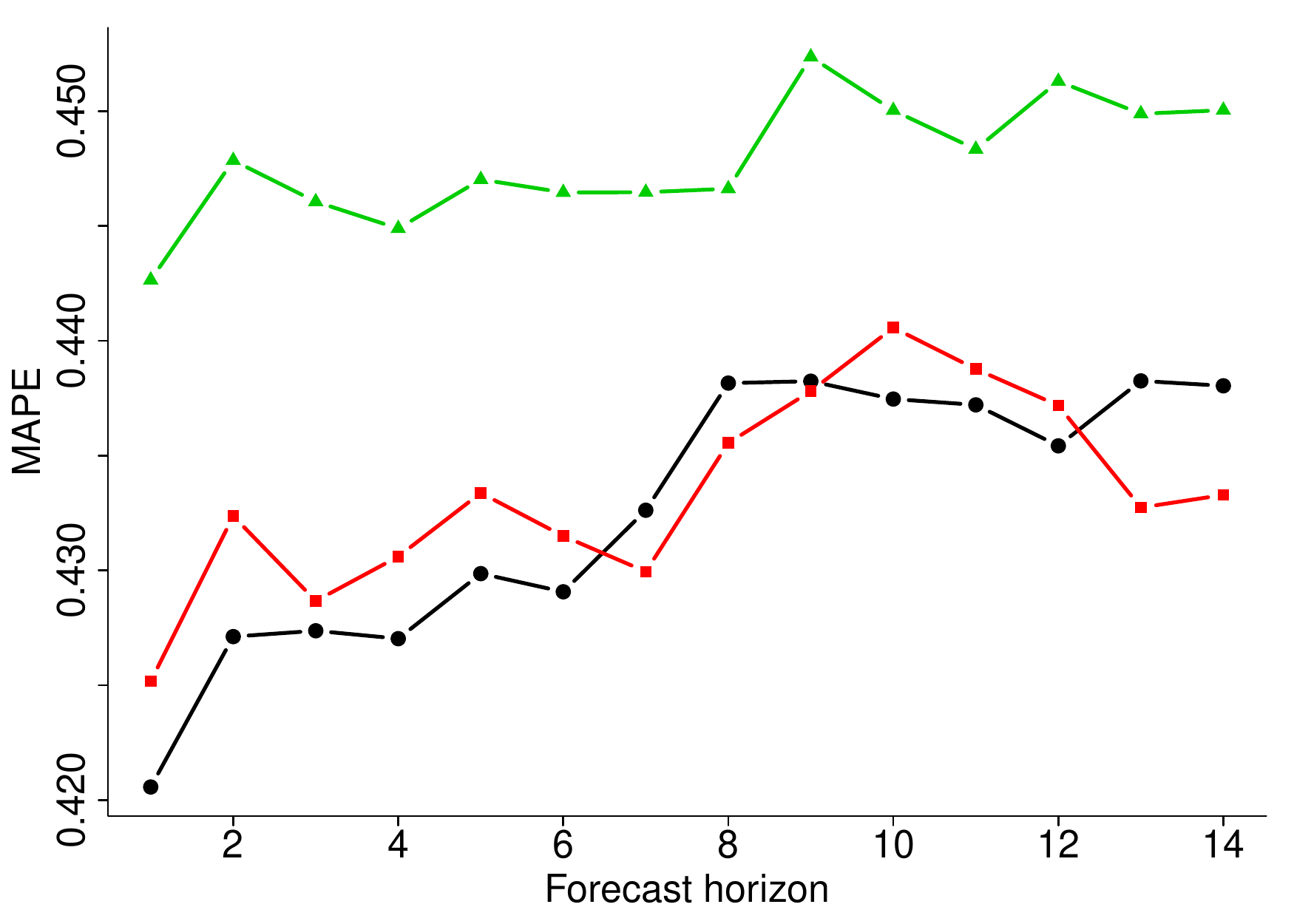} \\
 		\multicolumn{2}{c}{(i) Item A} \\
 		\includegraphics[width=.5\textwidth]{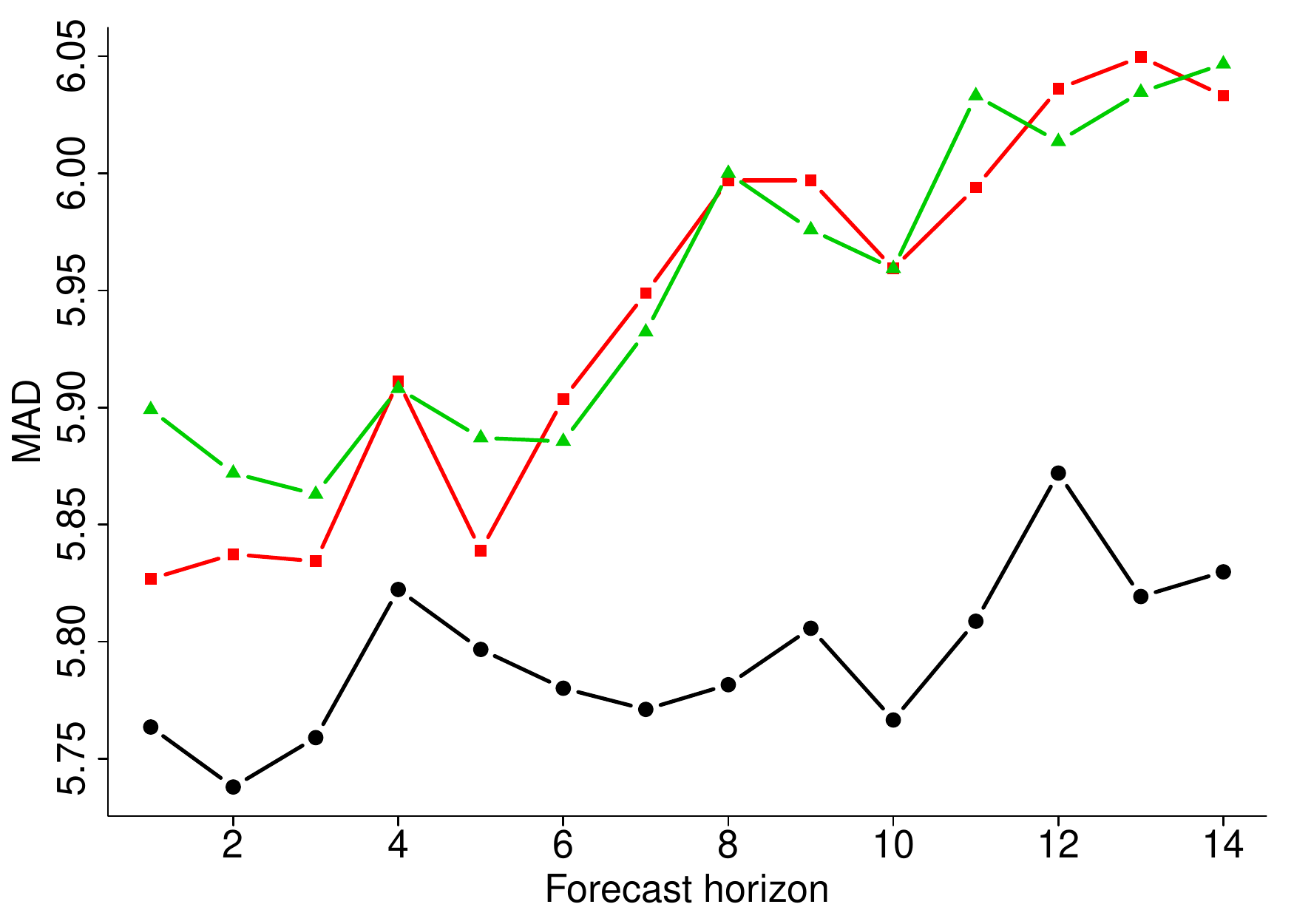} & 
 		\includegraphics[width=.5\textwidth]{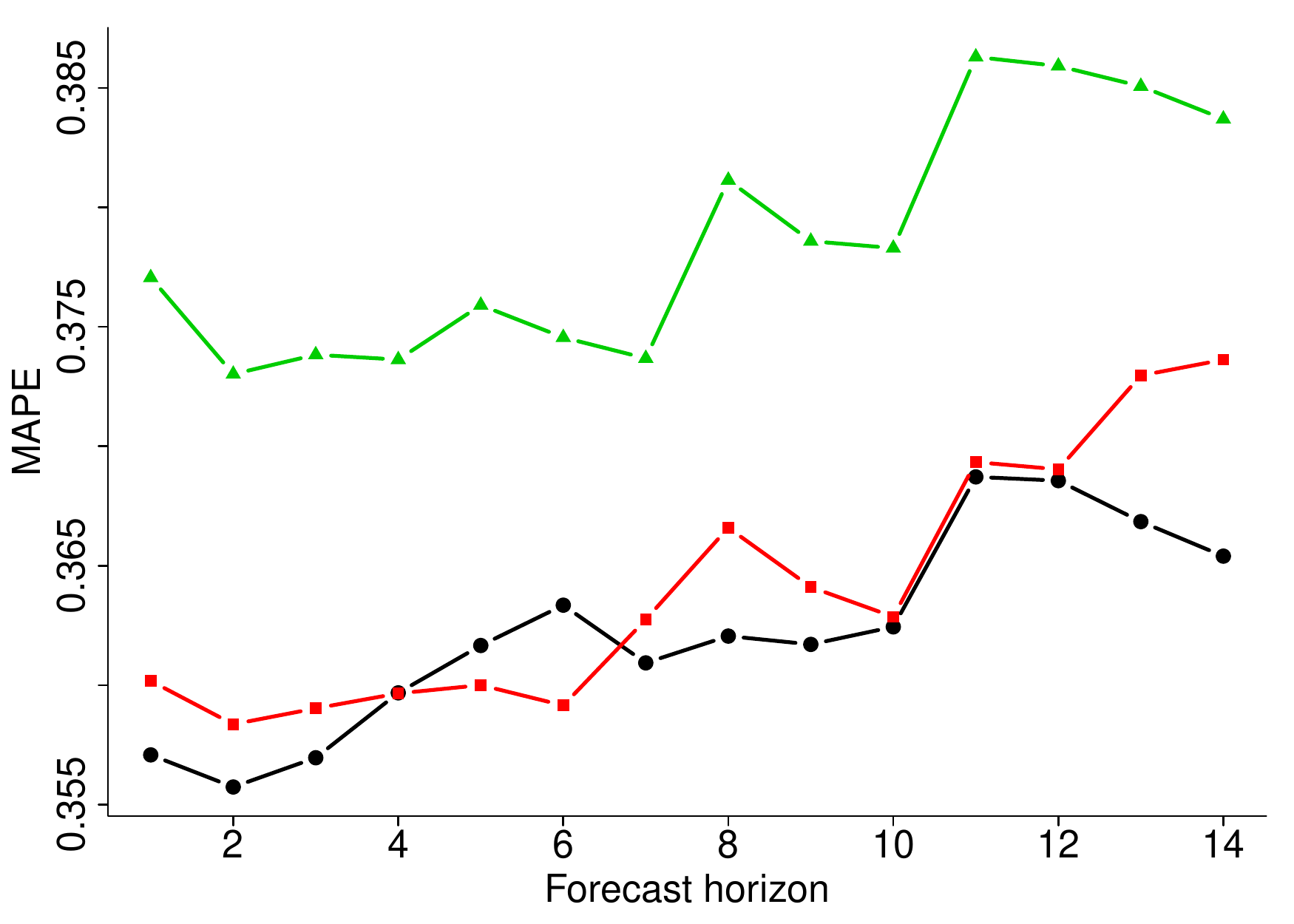} \\
 		\multicolumn{2}{c}{(ii) Item B} \\
 		\includegraphics[width=.5\textwidth]{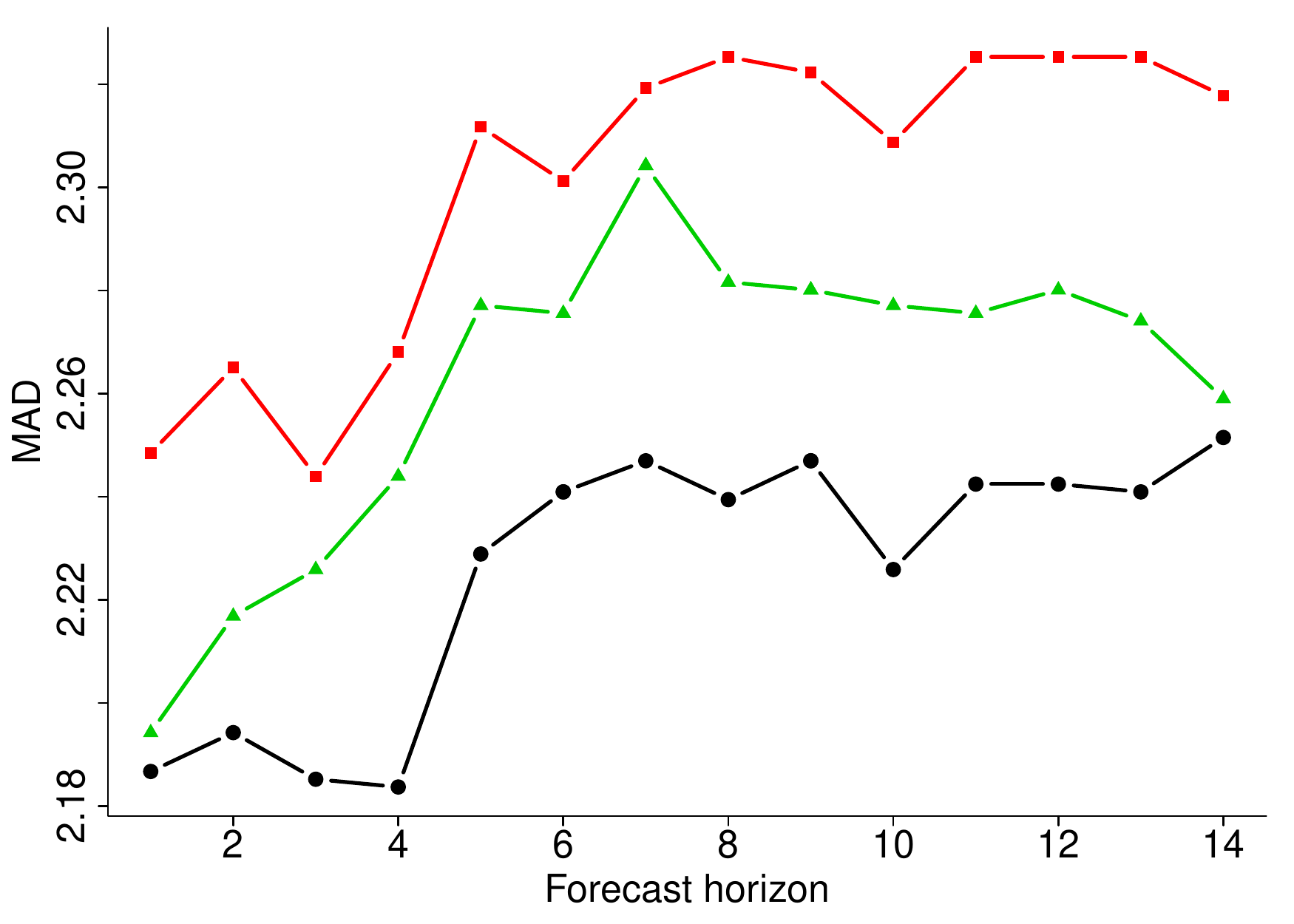} & 
 		\includegraphics[width=.5\textwidth]{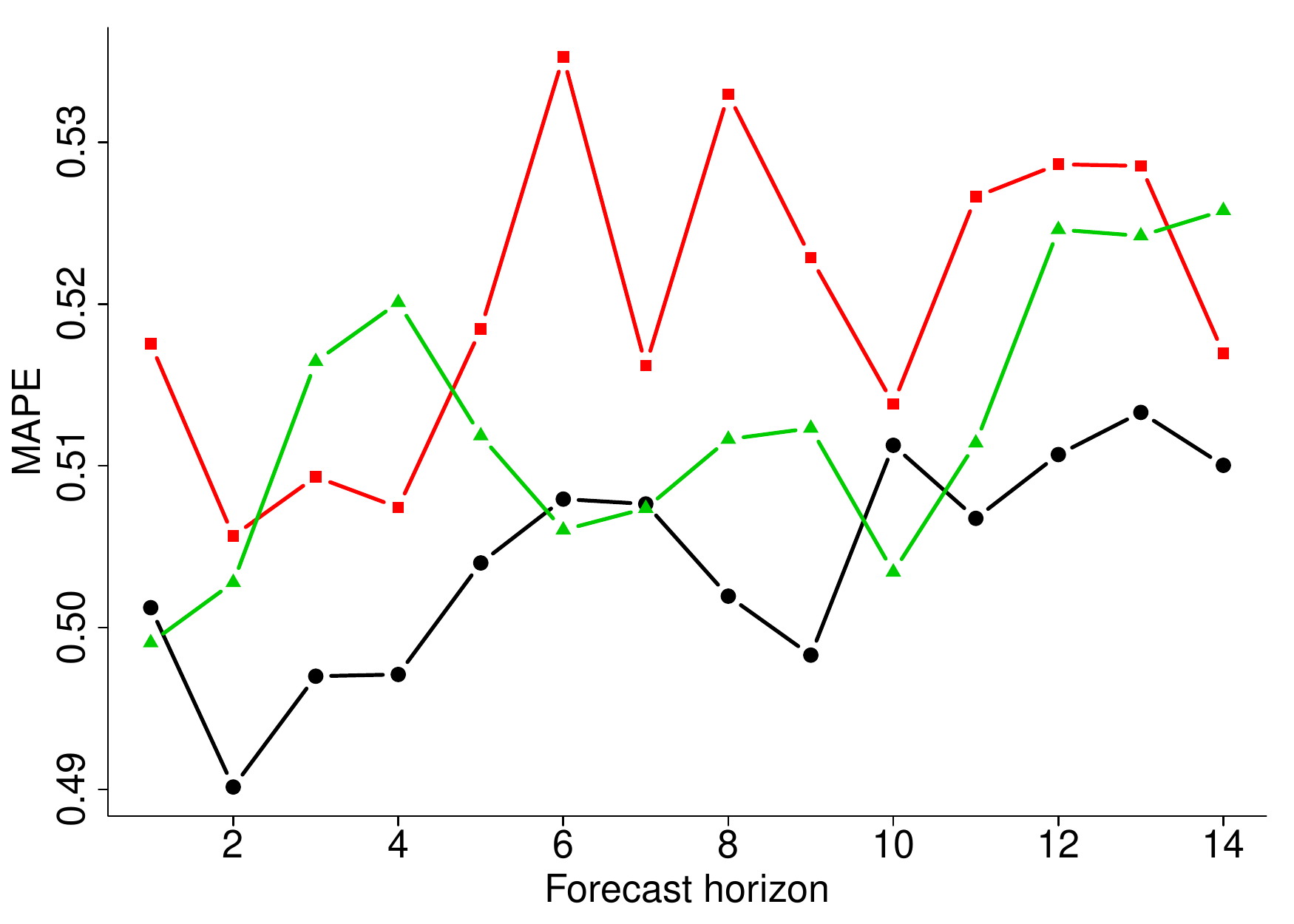}  \\ 
 		\multicolumn{2}{c}{(iii) Item C} \\
 	\end{tabular}
 	\caption{Mean absolute deviation (MAD: left) and mean absolute percentage error (MAPE: right) vs forecast horizon (days) for items A (top), B (middle), and C (bottom) from the multi-scale DBCM (black circles), independent DBCM (red squares), and independent DCMM (green triangles).}\label{fig:errors}
 \end{figure}

Figure~\ref{fig:distr_eval} (left column) displays coverage of the forecast distributions for $1$, $7$, and $14$-day ahead forecasts for each item. These plots show the empirical coverage obtained over the 322-day forecast period for predictive credible intervals (HPD - highest posterior density) of different percentages. Ideally, the empirical coverage of our credible intervals is close to the nominal level, resulting in coverage close to the $45-$degree line. For item A, the empirical coverage of credible intervals is close to the nominal coverage, although there is some evidence of slight under-coverage. For example, empirical coverage of $1$-step ahead $65\%$ credible intervals is about $60\%$. For item B, the empirical coverage of credible intervals is close to the nominal coverage. For $5\%$ and $20\%$ credible intervals, there is of slight over-coverage and for $65\%$ and $80\%$ intervals, there is slight under-coverage. For item C, forecast intervals have slight over-coverage. For example, the empirical coverage of $1$-day ahead $65\%$ intervals is about $71\%$. 

Figure~\ref{fig:distr_eval} (right column) displays randomized probabilistic integral transform (PIT; \citealp{kolassa2016evaluating}) values. If count valued data $y$ is forecast with predictive c.d.f., $P(\cdot)$, define $P(-1) = 0$ and draw a random quantity $p_y \sim U(P(y-1), P(y))$ given the observed value of $y$. Over repeat forecasts, an ideal model would generate values of $p_y$ that are approximately uniformly distributed. Figure~\ref{fig:distr_eval} plots  ordered randomized PIT values for $\seq{1}{14}$-day ahead forecasts versus uniform quantiles. For item A, the  values appear relatively uniform. Slight dips below the $45$-degree line could be random variation, or   may indicate that the lower tail of the forecast distribution is too light. For item B,   randomized PIT values appear to closely reflect uniform quantiles. For item C,   randomized PIT values are close to uniformity;  there are small dips below the $45-$degree line that could reflect random variability, or slightly underweight lower tails of forecast distributions. 

\subsubsection{Point Forecasts} 

Error metrics for selected point forecasts are shown in Figure~\ref{fig:errors}. We focus on two standard point forecast metrics, the mean absolute deviation (MAD) and the mean absolute percentage error (MAPE). Metrics are specific to a chosen lead-time $k>0$. For a series $y_t$, denote by $f_{t+k}$ a forecast of $y_{t+k}$ made at time $t$. MAD is the time average of the absolute deviation, $|y_{t+k} - f_{t+k}|$, and the optimal point forecast is the $k$-step ahead predictive median. MAPE, a common error metric in demand forecasting, is simply the time average of $|y_{t+k} - f_{t+k}|/y_{t+k}$, and the optimal point forecast is the $k$-step predictive $(-1)$-median. The $(-1)$-median of a distribution $f(y)$ is the median of $g(y)$ where $g \propto f(y)/y$. When evaluating the chosen error metrics, we use the corresponding optimal point forecast from each model. For each metric, we evaluate the error across $\seq{1}{14}$ days ahead on each day. The benchmark DCMM and both DBCM models (multi-scale and independent) are evaluated across a range of DCMM random effect discount factors, $\rho \in \{.2, .4, .6, .8, 1\}$. The accuracy of forecasting under each random effect may depend on the forecasting horizon, so we report only the lowest error across each of the five discount factors. Figure~\ref{fig:errors} displays the error from the best baseline DCMM, independent DBCM, and multi-scale DBCM across item, forecasting horizon, and metric.

\paragraph{Comparisons under MAD:} 
\begin{itemize}\itemsep=-3pt
	\item[A:] The multi-scale DBCM has the lowest MAD across the entire forecast horizon. Across the forecast horizon, the multi-scale DBCM has an average $1.4\%$ decrease in MAD compared to the DCMM. The multi-scale DBCM results in the largest percentage decreases in MAD for short- and mid-range forecasts of $1-3$ and $6-9$ days ahead. The independent DBCM and DCMM have similar MAD performance. 
	\item[B:] The multi-scale DBCM has the lowest MAD across the entire forecast horizon. Across the forecast horizon, the multi-scale DBCM has a average of a $2.6\%$ decrease in MAD compared to the DCMM. The largest percentage decreases in MAD occur for mid- to long-range forecasts of $7-14$ days ahead. The independent DBCM and DCMM have similar MAD performance. 
	\item[C:] The multi-scale DBCM has the lowest MAD across the entire forecast horizon. Across the forecast horizon, the multi-scale DBCM has a average of a $1.6\%$ decrease in MAD compared to the DCMM. The multi-scale DBCM has the largest percentage decrease in MAD in mid-range forecasts of $3, 4, 5, 7, 8$, and $10$-days ahead. The DCMM has lower MAD than the independent DBCM across the entire forecasting horizon.
\end{itemize} 

\paragraph{Comparisons under MAPE:} 
\begin{itemize}\itemsep=-3pt
	\item[A:] The multi-scale and independent DBCMs have lower MAPE across the entire forecast horizon. Across the forecast horizon, the multi-scale DBCM had an average decrease in MAPE of $3.4\%$ compared to the DCMM. The largest percentage drops in MAPE occurred for shorter-term forests from $1-6$ days ahead. 
	\item[B:] The multi-scale and independent DBCMs have lower MAPE across the entire forecast horizon. Across the forecast horizon, the multi-scale DBCM had an average decrease in MAPE of $4.3\%$ compared to the DCMM. The largest percentage drops in MAPE occurred sporadically when forecasting $1, 2, 8, 11, 13$, and $14$-days ahead. 
	\item[C:] The multi-scale DBCM has the lowest MAPE for 10 of 14 forecast horizons. Across the entire forecast horizon, the multi-scale DBCM had an average decrease in of $1.6\%$ compared to the DCMM. The largest improvements in MAPE occurred sporadically when forecasting $3, 4, 9$, and $14$-days ahead. The DCMM has lower MAPE than the independent DBCM for 11 out of 14 forecast horizons. 
\end{itemize} 

\subsubsection{Forecasting and Impact of Excess}
It is also of interest to exemplify the dissection of forecasts based on the binary cascade excess distribution, and explore the impact on forecast uncertainties in particular. From the simulation-based DBCM joint forecast distributions we can trivially extract predicted probabilities of no excess on a future day-- the probability than none of the transactions on that day sell more than the specified $d$ items.  At the store level, this is potentially useful additional summary information in its own right.  Further,  looking at the sales forecast distributions conditional on no excess baskets on a particular day provides insights into the impact-- on both forecast level and uncertainties-- of the excess component of the model. 

\begin{figure}[hp]
 	\centering
 	\begin{tabular}{cc}
 		\includegraphics[width=.5\textwidth]{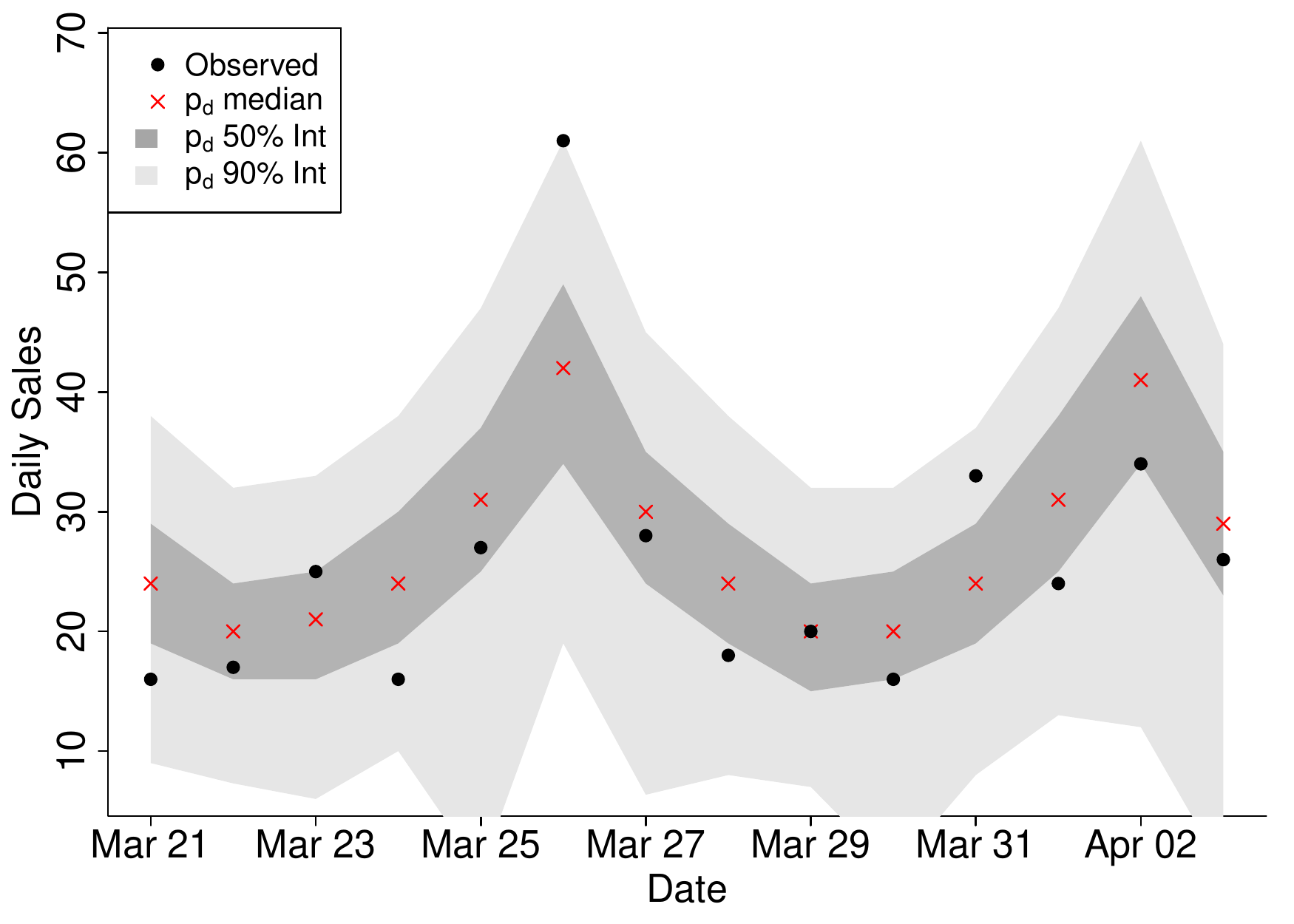} & 
 		\includegraphics[width=.5\textwidth]{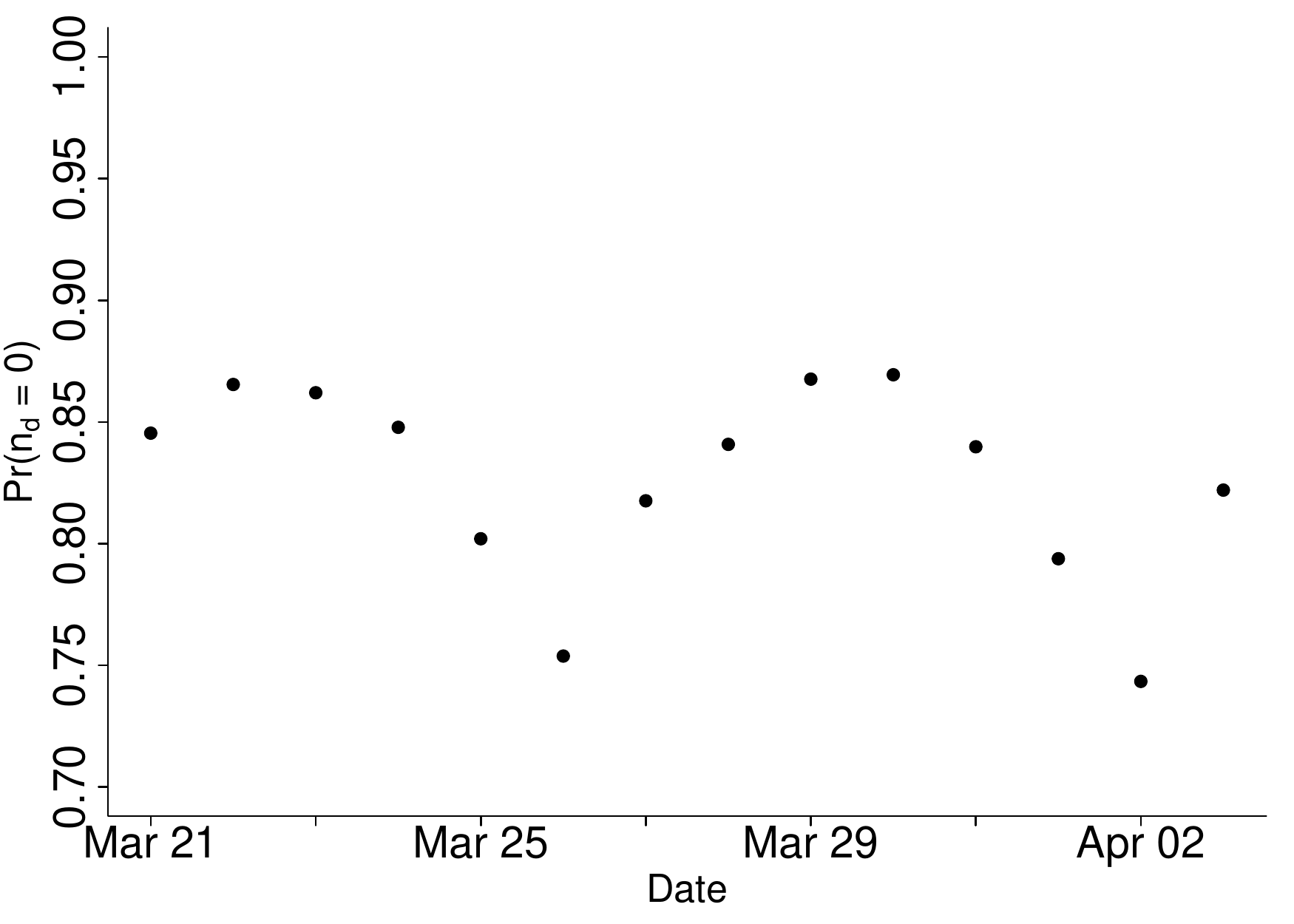} \\
 		\multicolumn{2}{c}{(i) Item A} \\
 		\includegraphics[width=.5\textwidth]{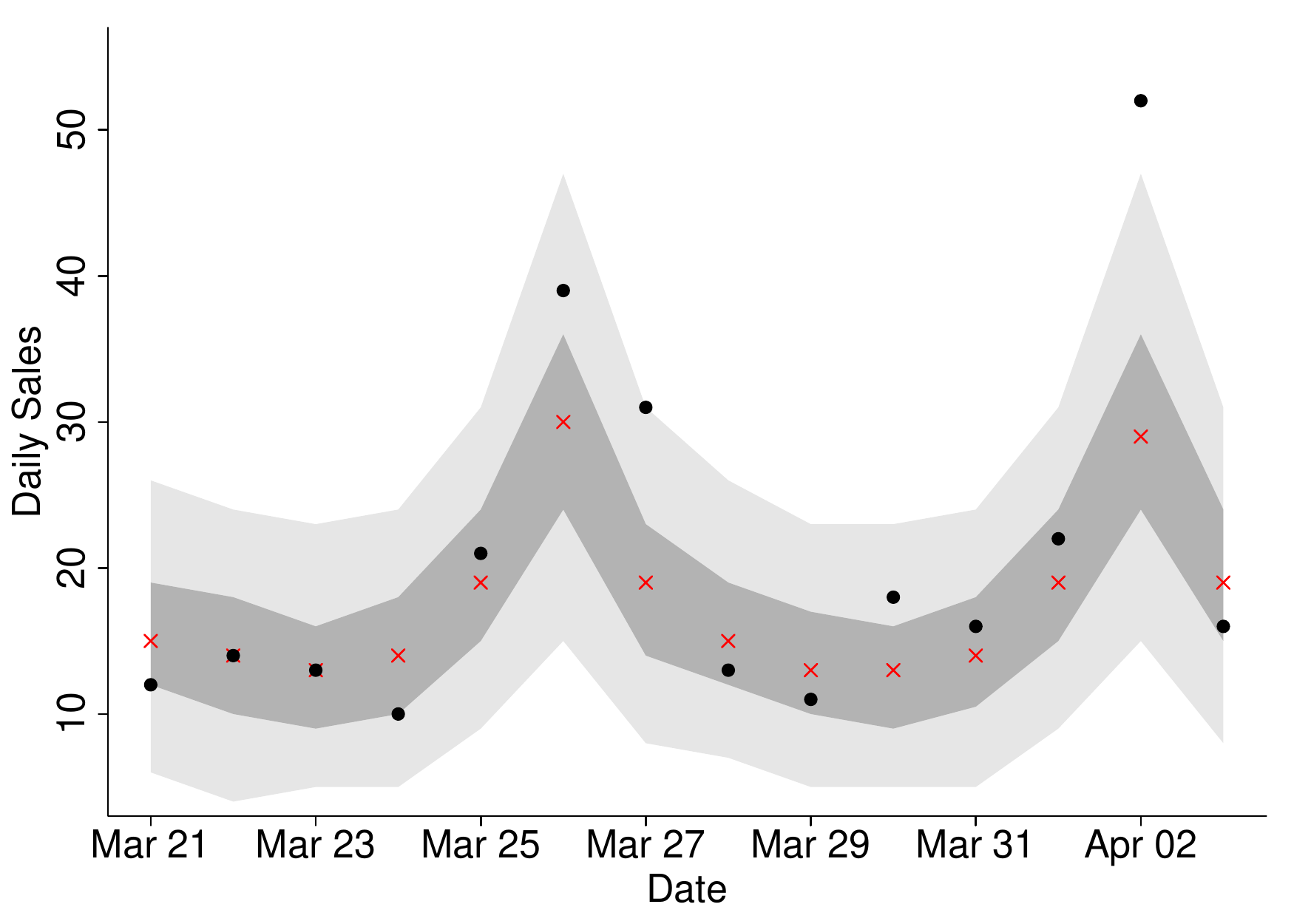} & 
 		\includegraphics[width=.5\textwidth]{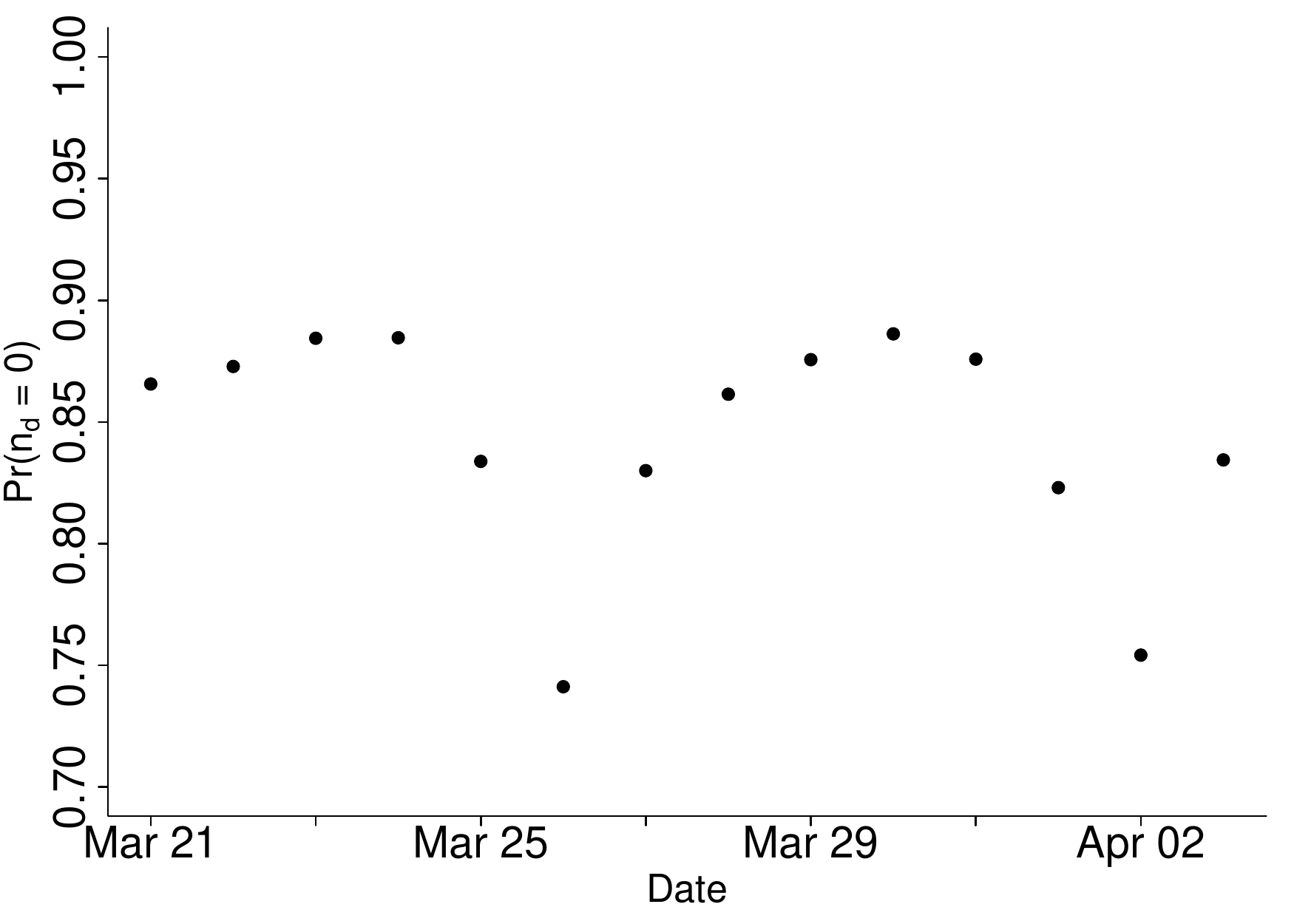} \\
 		\multicolumn{2}{c}{(ii) Item B} \\
 		\includegraphics[width=.5\textwidth]{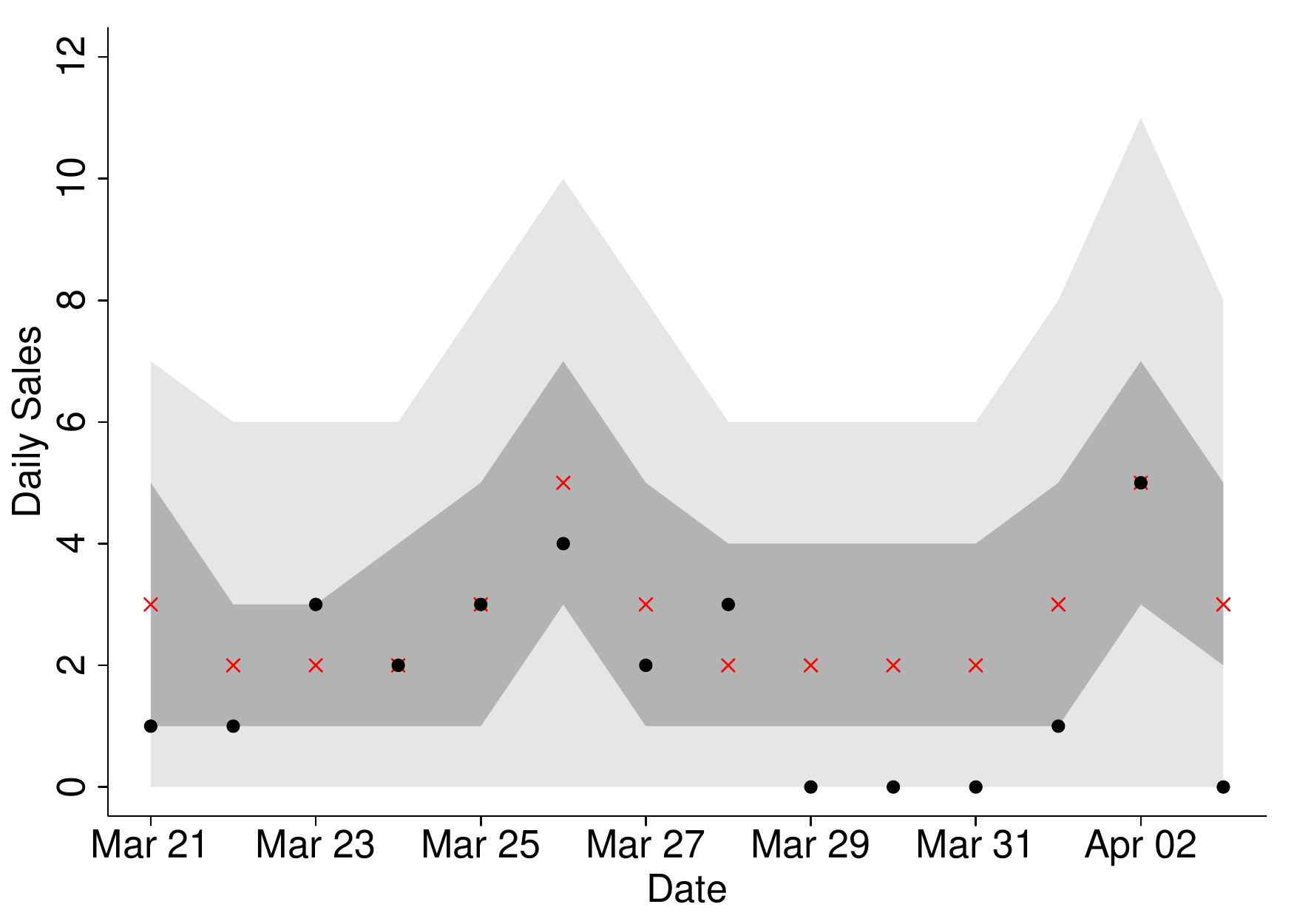} & 
 		\includegraphics[width=.5\textwidth]{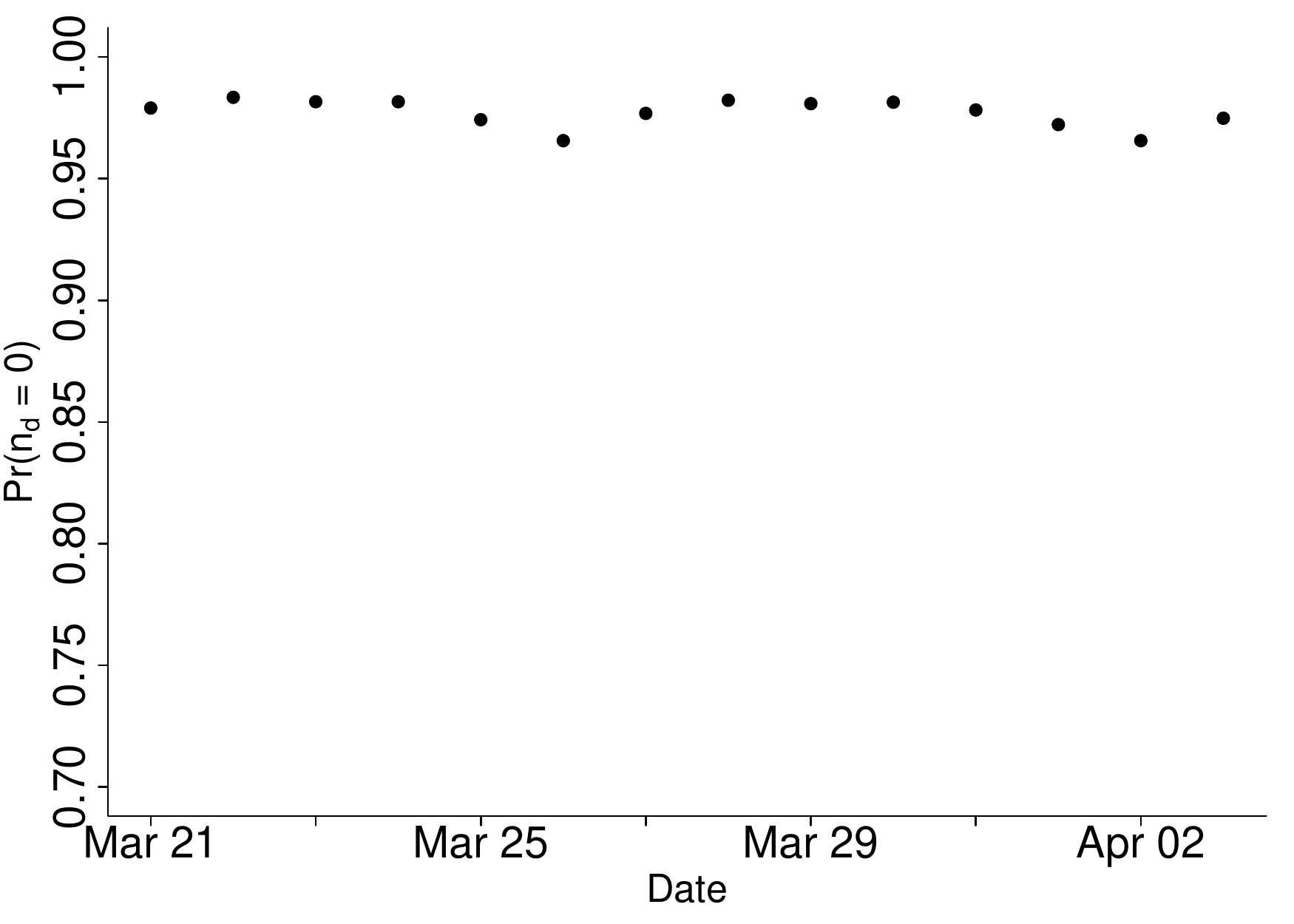} \\
 		\multicolumn{2}{c}{(iii) Item C} \\
 	\end{tabular}
 	\caption{1-14 day forecasts made on on Mar 20th 2017.  Joint forecast trajectories conditional on no excess baskets (left),  with details as in 
 	unconditional trajectories in Figure~\ref{fig:trajectories}, and of corresponding probabilities of no excess (right).}\label{fig:condtlforecastexample}
 \end{figure}

One selected example is summarized in Figure~\ref{fig:condtlforecastexample} using 1-14 day forecasts for each item made at the earlier selected date of Mar 20th 2017. The figure shows the trajectories of joint forecast  distributions over the next 14 days now conditional on no excess (i.e., conditional on predicted $n_{d,t+k}=0$ for $k=1:14$ where $t$ indexes Mar 20th 2017). These figures have the same format as those for the full unconditional forecasts shown in Figure~\ref{fig:trajectories}. Small differences can be seen, with the conditional forecast distributions naturally favoring slightly lower values while being less diffuse; this is also naturally more pronounced for higher levels of sales such as for item A. Figure~\ref{fig:condtlforecastexample} also displays trajectories of the predictive probabilities of no excess over the next 14 days, naturally indicating higher probabilities for the lower levels of sales exhibited by item C.

\section{Summary Comments}
\label{sec:conc} 

Motivated by an application to product demand forecasting, and enabled by the availability of rich point-of-sale data, we have introduced a novel framework for Bayesian state space modeling of heterogeneous transactions-sales time series. This work stems from the recognition that variability seen in high frequency sales arises from the compounding effect of variability in the number of transactions as well as the number of sales-per-transaction. The dynamic binary cascade model builds upon prior approaches to univariate count time series, notably the DCMM of~\cite\dcmm. The initial stage in this framework involves adapting the DCMM to model transactions rather than sales. Given the reduced variability of transactions relative to sales, this is a promising application in which the DCMM may improve forecasting accuracy. 

Application of the DCMM to transactions of related items offers an opportunity to integrate information across series through a multi-scale, multivariate dynamic factor model. Coupled with the DCMM on transactions, the binary cascade concept involves a sequence of Bayesian models to predict the number of units sold per transaction. The motivation behind this binary cascade is that the appropriate way to forecast rare events is through a sequence of conditional probabilities which define chances of outcomes of increasingly higher-- and rarer-- sales per transaction.
The final stage of the DBCM framework is the choice of excess distribution -- leaving it unspecified or choosing a specific form. Leaving the excess distribution unspecified avoids the difficult task of fitting the long tail of the sales-per-transaction distribution, however, this approach limits the conclusions we can present about the forecast distribution. We also present a logical nonparametric choice for the excess distribution which involves bootstrapping from the empirical excess distribution. 

In addition to the incorporation of covariates into the binary and Poisson DGLM components of the DCMM, the DBCM framework extends the hierarchical decomposition further by incorporating covariates into the cascade of binomial logistic DGLMs. This allows incorporation of complex price/promotion effects which may impact the overall traffic in the store, the probability that a customer makes a purchase, and the number of units purchased given that a transaction occurs. 
The Bayesian framework used for the DBCM allows direct/forward simulation of multi-step ahead predictions, enabling trivial computation of forecast summaries of interest. 
Selected examples of sales forecasting show the promise for forecast improvement of the DBCM across demand sizes, error metrics, and forecast horizon, emphasizing assessment of probabilistic forecasting accuracy in multiple metrics as well as via standard point forecast summaries.  

Future studies will explore the benefits and drawbacks of the DBCM approach across very large numbers of items and across multiple retail outlets.  One specific applied component of the models open to further development is the integration of additional, feed-forward information about promotions at the item level. This of particular interest in connection with forecasting infrequent higher basket sizes based on, for example, \lq\lq buy 1, get 1 free'' types of promotion. Such information can be incorporated in modified models of the excess distribution in a number of ways that should yield practical forecast improvements in such cases. Finally, 
in addition to contributing advances in dynamic model-based forecasting for consumer sales, the new class of DBCMs should be of interest in other areas involving multiple heterogeneous time series  of 
non-negative integers.  This includes areas such as marketing and modeling consumer behavior in other contexts where counts arise from underlying compound processes such as in forecasting visitors to different tourist sites by first forecasting numbers of cars, and then number of passengers per car. 
 
\newpage

\appendix
\section{Appendix: Technical Details of Dynamic Models\label{app:models} }

\subsection{General Setting and Notation of DGLMs} 
The new class of dynamic transaction-sales models has components involving binary, binomial and Poisson dynamic generalized linear models with sampling distributions and regression forms in~\eqnotwo{dccmstates}{DBCM}.  Models for higher-level aggregate demand in the multi-scale context may adopt
traditional normal dynamic linear models for log data.   These are   standard state-space models and full background and details of analyses can be found in, for example~\cite{west1997book} and~\cite{Prado2010}.   Data are modeled on their natural scale, and components of state-space models such as levels, trends, seasonality, and regression components are easily interpretable. Time-varying state vectors  allow models to adapt over time and accommodate unpredictable changes.   Sequential learning and forecasting involves state vectors evolving in time with information changing via prior-posterior updates at each time. Models are open to  incorporation of expert information or interventions   at any time  via modifications of  priors over state parameters.

Key summaries, including model structure, forward filtering and forecasting aspects, are given here. 
This is presented for a DGLM in a general setting with the following notation and structure.
\begin{itemize} \itemsep-3pt
\item  Over time $t,$  scalar $y_t$ denotes the time series of interest, whether it be continuous, binary, or non-negative count.   
\item At any time $t$ having observed $y_{1:t},$ available information is denoted-- and sequentially updated -- by 
$\cD_t = \{ y_t, \cD_{t-1},\cI_{t-1} \}$ where $\cI_{t-1}$ represents any relevant additional information becoming available at time $t-1$ in addition to past data (such as information used to define interventions in the model).  For any  vector of time indices $\seq{t+1}{t+k}$ for $k>0,$ forecasting $y_{\seq{t+1}{t+k}}$ at time $t$ is based on the information set $\{ \cD_t,\cI_t \}. $ 
\item   $\F_t,\btheta_t$ are the time $t$ dynamic regression vector and state vector, respectively. 
\item The state-space structure is
\begin{equation} \label{eq:DGLMregnevo}
\lambda_t = \F'_t \btheta_t\quad\textrm{where}\quad \btheta_t = \G_t \btheta_{t-1} + \bomega_t\quad\textrm{and}\quad \bomega_t \sim (\bzero,\W_t)
\end{equation} 
where $\lambda_t$ is the linear predictor at time $t$. 
\item  This conditionally linear, Markov process model for $\btheta_t$ over time has known state evolution (or transition) matrix  
$\G_t$  and  stochastic innovation vector (or evolution \lq\lq noise'') $\bomega_t$.   
\item The $\bomega_t$ are conditionally independent and also independent of the current and past states, and have moments 
$\text{E}[\bomega_t|  \cD_{t-1},\cI_{t-1} ]=\bzero$ and $\text{V}[\bomega_t | \cD_{t-1},\cI_{t-1} ] = \W_t$,  known at time $t-1.$   
\end{itemize} 

\subsection{Sequential Learning \label{app:sequentiallearning} } 

\subsubsection{Non-normal cases \label{sec:DGLMappnonn}}
In a model with binomial or Poisson structure,   the linear predictor $\lambda_t$ is a one-to-one transformation of the natural parameter of the sampling distribution.  As in ~\eqnotwo{dccmstates}{DBCM} these involve logistic and log transforms.   
The one-step analysis over times $t-1$ to $t$ utilize constraints to conjugate priors/posteriors for the natural parameters to enable closed-form updating and access to relevant predictive distributions for forecasting.  The details are as follows~(e.g., ~\citealp{west1985dynamic};~\citealp{west1997book}~chapter~15;~\citealp{Prado2010}~section~4.4).
\begin{enumerate}[(a)] \itemsep=-3pt
\item  At $t-1$,  current information is summarized via the mean vector and variance matrix of the posterior for the current state vector, namely 
  $(\btheta_{t-1} \mid \cD_{t-1}, \cI_{t-1}) \sim [\m_{t-1}, \C_{t-1}]$. 
 \item \label{dglm:Prior} The implied $1-$step ahead prior moments for the time $t$ state vector are 
 $(\btheta_{t} \mid \cD_{t-1}, \cI_{t-1}) \sim [\a_{t}, \R_{t}]$ with $\a_t = \G_{t} \m_{t-1}$ and $\R_{t} = \G_{t} \C_{t-1} \G'_t + \W_t$. 
 \item  \label{dglm:VBprior}  The time $t$ prior is chosen to be of conjugate form with parameters defined by the prior moments of
 $\lambda_t$ implied by point \ref{dglm:Prior} above, i.e., the conjugate prior satisfies 
$$\text{E}[\lambda_t \mid \cD_{t-1}, \cI_{t-1}]  = f_t =\F'_t \a_t \quad\textrm{and}\quad   \text{V}[ \lambda_t  \mid \cD_{t-1}, \cI_{t-1} ]  = q_t = \F'_t \R_t \F_t.$$
\item \label{eq:expfamforecast} Forecasting $y_t$  $1-$step ahead uses the conjugacy-induced predictive distribution with p.d.f.   
$p(y_t \mid \cD_{t-1},\cI_{t-1}).$   In all conjugate models this is of known analytic form and can be simulated trivially.  

\item  \label{dglm:conjpost}
On observing $y_t,$ the posterior for $\lambda_t$ is implied by the conjugate  form posterior for the natural parameter. 
\item  \label{dglm:VBpost} Under this posterior,  mapping back to the linear predictor $\lambda_t = g(\eta_t)$  implies posterior mean and variance 
$g_t = \text{E}[\lambda_t \mid \cD_t ] $ and $p_t = \text{V}[\lambda_t \mid \cD_t ].$ 
\item \label{dglm:VBpostmoms}  Linear Bayes updating~\citep{GoldsteinWooff2007}  gives the  posterior mean vector and variance matrix in 
$ (\btheta_t \mid \cD_t ) \sim [\m_t, \C_t]$ as
$$  \m_t = \a_t +  \R_t \F_t (g_t - f_t) /q_t\quad\textrm{and}\quad  \C_t = \R_t - \R_t \F_t \F'_t \R'_t (1 - p_t/q_t)/q_t.$$
This completes the time $t-1$-to-$t$ evolve-predict-update cycle.
\end{enumerate} 
This general structure specializes in the binary, binomial and Poisson models  as follows.

\medskip\noindent{\bf\em Binomial logistic DGLM: }  Binary or binomial DGLMs are used for the DCMM model component for zero/non-zero transactions $z_t$ in \eqno{DCMM} and for each of the component models in the dynamic binary cascade  for sales per transaction $n_{r,t}$ in \eqno{DBCM}.  The binary case-- that of a Bernoulli DGLM-- is simply a special case of the binomial model summarized here. 

Here the series $y_t$ is conditionally binomial with, in a general notation, $y_t \sim Bin(h_t, \pi_t)$ where $h_t$ is the positive integer \lq\lq number of trials'' and the success probability $\pi_t$ relates to the linear predictor via $\lambda_t = \text{logit}(\pi_t)$.    The binary case has, of course, $h_t=1.$ 
The conjugate prior in step~(\ref{dglm:VBprior}) above is Beta, $\pi_t \sim Be(\alpha_t, \beta_t)$, with the hyper-parameters defining  $f_t = \psi(\alpha_t) - \psi(\beta_t)$ and $q_t = \psi'(\alpha_t) + \psi'(\beta_t)$, where $\psi(\cdot)$ and $\psi'(\cdot)$ are the digamma and trigamma functions, respectively. The values $(\alpha_t,\beta_t)$ can be trivially computed from $(f_t,q_t)$ via iterative numerical solution based on standard Newton-Raphson.  The $1-$step ahead forecast  is Beta-Bernoulli with   $(y_t \mid \cD_{t-1},\cI_{t-1}) \sim BBer(h_t, \alpha_t, \beta_t).$    The conjugate posterior in step~(\ref{dglm:conjpost}) above is $\pi_t \sim Be(\alpha_t+y_t, \beta_t+h_t-y_t)$.  The updated moments of the linear predictor in step~\ref{dglm:VBpost} above are then  trivially computed via the equations $g_t = \psi(\alpha_t+y_t) - \psi(\beta_t+h_t-z_t)$ and $p_t = \psi'(\alpha_t+y_t) + \psi'(\beta_t+h_t-z_t).$

\medskip\noindent{\bf\em Poisson loglinear DGLM:}  In the DCMM model component we have shifted Poisson data based on the Poisson DGLM.  The general Poisson analysis is trivially applied to the time series shifted by 1 unit. 

In the general setting,  $y_t \sim Po(\mu_t)$ with $ \lambda= \log(\mu_t)$.     The conjugate prior
in step~(\ref{dglm:VBprior}) above is Gamma, $\mu_t \sim Ga(\alpha_t, \beta_t)$, with the hyper-parameters defining  $f_t = \psi(\alpha_t) - \log(\beta_t)$ and $q_t = \psi'(\alpha_t).$  The values $(\alpha_t,\beta_t)$ can be trivially computed from $(f_t,q_t)$ via iterative numerical solution based on standard Newton-Raphson.  The $1-$step ahead forecast  is negative binomial, $(y_t \mid \cD_{t-1},\cI_{t-1}) \sim Nb(\alpha_t, \beta_t/(1 + \beta_t)$. 
 The conjugate posterior in step~(\ref{dglm:conjpost}) above is $\mu_t \sim Ga(\alpha_t+y_t, \beta_t+1).$
 The updated moments of the linear predictor in step~(\ref{dglm:VBpost}) above then are  trivially computed via the equations $g_t = \psi(\alpha_t+y_t) - \log(\beta_t+1)$ and $p_t = \psi'(\alpha_t+y_t).$

\subsubsection{Normal cases}

When $y_t$ is conditionally normal, the DGLM reduces to a conditionally normal DLM.   This is of relevance to count time series in case of large counts where a log transform-- for example-- of the count series can often be well-modeled using a normal DLM as an approximation. This also allows for inclusion of volatility via a time-varying conditional variance.    

In the general setting,  $y_t \sim N(\mu_t, v_t)$ with $ \lambda_t=  \mu_t$ defining the dynamic regression and $v_t$ a  potentially time-varying variance.  Consider first the case of $v_t$ known.  The conjugate prior in step~(\ref{dglm:VBprior}) of Section~\ref{sec:DGLMappnonn} above is normal as is the $1-$step ahead forecast distribution and the implied posterior for $\mu_t.$  The prior to posterior updating in step~(\ref{dglm:VBpostmoms}) reduces to a standard Kalman filter update.  When embedded in the DLM, the   additional  assumption that the evolution noise terms $\bomega_t$ in \eqno{DGLMregnevo}
are also normal implies that DGLM evolution/updating equations are exact in this special case.  However,  for most practical applications it is relevant to also estimate the conditional variances  $v_t = 1/\phi_t.$ The simplest and most widely-used extension is that based on a standard Beta-Gamma stochastic volatility model for $\phi_t$ which,  is analytically tractable.  The resulting theory is then based on normal/inverse gamma prior and posterior distributions for $(\mu_t,v_t). $ Details of the resulting modifications to forward filtering and forecasting analysis are very standard~(\citealp{west1997book}~chapter~4 and section~10.8;~\citealp{Prado2010}~section~4.3).

\subsection{Discount Factor for Evolution Variance Matrices \label{app:discountfactors} } 

Values of variance matrices $\W_t$ in  \eqno{DGLMregnevo} use  component discounting~\citep[][chapter 6]{west1997book}. In practical models the state vector is partitioned into components representing different 
explanatory effects, such as trends (e.g., local level, local gradient), seasonality (time-varying seasonal factors or Fourier coefficients) and coefficients of predictor variables. Then, for some integer $J$ we have $\btheta_t' = (\btheta_{t1}',\ldots,\btheta_{tJ}')$. It is then natural to define $\W_t$ to represent potentially differing degrees of stochastic variation in these components and this is enabled using separate discount factors $\delta_1,\ldots,\delta_J,$ where each $\delta_j\in (0,1].$ A high discount factor implies a low level of stochastic change in the corresponding elements of the state vector, and vice-versa (with $\delta_j=1$ implying no stochastic noise at all-- obviously desirable but rarely practically relevant).      

From Appendix~\ref{app:sequentiallearning} part \ref{dglm:Prior} above, the time $t-1$ prior variance matrix of $ \G_t\btheta_{t-1}$  is $\P_{t} = \G_{t} \C_{t-1} \G'_t;$  this represents information levels about the state vector following the deterministic evolution via $\G_t$ but before the impact of the  evolution noise that then simply adds $\W_t.$   Write 
$\P_{tj}$ for the diagonal block of $\P_t$ corresponding to state subvector $\btheta_{tj}$ and set 
$$\W_t = \text{block diag}[ \P_{t1}(1-\delta_1)/\delta_1 , \ldots, \P_{tJ}(1-\delta_J)/\delta_J].$$ Then the implied prior variance matrix of $\btheta_t$ following the evolution has corresponding diagonal block elements $\R_{tj} = \P_{tj}/\delta_j$ while maintaining off-diagonal blocks from $\P_t.$ 
Thus, the stochastic part of the evolution increases uncertainties about state vector elements in each subvector $j$ by $100(1-\delta_j)/\delta_j\%,$ 
maintains the correlations in $\P_{tj}$ for state elements within the subvector $j,$ while reduces cross-correlations between state vector elements in differing subvectors. In practice, high values of the $\delta_j$ are desirable and typical applications use values in the range $0.97-0.99$ with, generally, robustness in terms of forecasting performance with respect to values in the range. Evaluation of forecast metrics on training data using different choices of discount factors is a basic strategy in model building and tuning.

\subsection{Dynamic Random Effects Extensions of State-Space Model Components}

One extension of the traditional  DGLMs used as a key model component that is very relevant to transaction forecasting 
involves the introduction of additional, time specific random effects in the state vector. This can be used in binary and shifted Poisson components as it is generally applicable to any DGLM,  but here is of main interest and potential importance in the conditional Poisson component as it has the ability to capture additional variation in count levels beyond that predicted by a core state-space model.  
This was introduced in~\cite\dcmm\ in DCMMs for sales forecasting. In the current work, while the new coupled transactions-sales model already explicitly dissects observed variations in sales by accounting for heterogeneity of basket size across transactions, for many items there will typically still be a need to represent and estimate additional day-specific, unpredictable variation that goes beyond that captured by the specific model.  

We summarize this here in the context of the shifted Poisson DGLM  for transactions.  Use the modified notation of 
$\F_{t,0},\btheta_{t,0}$ and $\lambda_{t,0} = \F'_{t,0} \btheta_{t,0}$ for the dynamic regression vector, state vector and linear predictor in a specified \lq\lq baseline''  model.  Then define a random effects extended model to have
state vector $\btheta_t =  (\zeta_{t}, \btheta_{t,0}')'$ and regression vector $\F_t = (1, \F_{t,0}')'$ where $\zeta_t$ is a series of independent, zero-mean random effects that are also independent of the current and past baseline state vectors. The implied linear predictor $\lambda_t$ and resulting Poisson mean $\mu_t$ are 
given by $\log(mu_t) = \lambda_t = \lambda_{t,0} + \zeta_t;$ so  the $\zeta_t$ provide additional, day-specific \lq\lq shocks'' to latent transaction rates, separately from the changes inferred by the predictive baseline model.  The model extension uses a random effects discount factor $\rho$, $(0<\rho\le 1),$ to define levels of random effects variability. This tuning parameter is used as follows. At  time 
$t-1$ prior uncertainty about the baseline state vector $\R_{t,0} = \textrm{V}[\btheta_{t,0} |  \cD_{t-1},\cI_{t-1}] $  implies
$q_{t,0} \equiv \text{V}[\lambda_{t,0}| \cD_{t-1},\cI_{t-1}] = \F_{t,0}'\R_{t,0}\F_{t,0}.$  The model sets 
$ v_t = \text{V}[\zeta_t| \cD_{t-1},\cI_{t-1}] =  q_{t,0}  (1-\rho)/\rho$ 
The baseline Poisson DGLM arises as the special case $\rho=1$ while a smaller value of $\rho$ induces a higher level of time $t$-specific variation, implying increased dispersion of forecast distributions.  
The DGLM analysis of Section~\ref{sec:DGLMappnonn} applies with a trivial extension to technicalities: the effective  result is that 
the prior variance $q_t$ of the linear predictor-- in Appendix~\ref{app:sequentiallearning} part \ref{dglm:Prior} above-- is modified 
to $q_t =  q_{t,0} + v_t = q_{t,0}/\rho.$  This makes clear that the discount factor $\rho$ defines an inflation of prediction variance 
relative to that of the baseline model.

\section*{Acknowledgments}
The research reported here was partly supported by $84.51^\circ$. 
We acknowledge discussions and data development with Xiaojie Zhou and others in the research team at  $84.51^\circ$.
 Any opinions, findings and conclusions or recommendations expressed in this paper do 
 not necessarily reflect the views of $84.51^\circ$.

%
\bibliographystyle{elsarticle-harv}
\bibliography{DBCM}

\begin{thebibliography}{21}
\expandafter\ifx\csname natexlab\endcsname\relax\def\natexlab#1{#1}\fi
\expandafter\ifx\csname url\endcsname\relax
  \def\url#1{\texttt{#1}}\fi
\expandafter\ifx\csname urlprefix\endcsname\relax\def\urlprefix{URL }\fi

\bibitem[{Aktekin et~al.(2018)Aktekin, Polson, and Soyer}]{Soyer2018}
Aktekin, T., Polson, N.~G., Soyer, R., 2018. Sequential {B}ayesian analysis of
  multivariate count data. Bayesian Analysis 13, 385--409.

\bibitem[{Arunraj and Ahrens(2015)}]{arunraj2015}
Arunraj, N.~S., Ahrens, D., 2015. A hybrid seasonal autoregressive integrated
  moving average and quantile regression for daily food sales forecasting.
  International Journal of Production Economics 170, 321--335.

\bibitem[{Berry and West(2018)}]{BerryWest2018dcmm}
Berry, L., West, M., 2018. Bayesian forecasting of many count-valued time
  series. Submitted for publication.\,ArXiv:1805.05232.

\bibitem[{Chen and Lee(2017)}]{chen2017}
Chen, C. W.~S., Lee, S., 2017. Bayesian causality test for integer-valued time
  series models with applications to climate and crime data. Journal of the
  Royal of Statistical Society (Series C: Applied Statistics) 66, 797--814.

\bibitem[{Chen et~al.(2016)Chen, So, Li, and Sriboonchitta}]{chen2016}
Chen, C. W.~S., So, M. K.~P., Li, J., Sriboonchitta, S., 2016. Autoregressive
  conditional negative binomial model applied to over-dispersed time series of
  counts. Statistical Methodology 31, 73--90.

\bibitem[{Croston(1972)}]{croston1972forecasting}
Croston, J.~D., 1972. Forecasting and stock control for intermittent demands.
  Operational Research Quarterly (1970-1977) 23~(3), 289--303.

\bibitem[{Goldstein and Wooff(2007)}]{GoldsteinWooff2007}
Goldstein, M., Wooff, D.~A., 2007. Bayes Linear Statistics: Theory and Methods.
  Chichester: John Wiley.

\bibitem[{Kolassa(2016)}]{kolassa2016evaluating}
Kolassa, S., 2016. Evaluating predictive count data distributions in retail
  sales forecasting. International Journal of Forecasting 32, 788--803.

\bibitem[{Kolassa(2018)}]{kolassa2018}
Kolassa, S., 2018. Commentary on retail forecasting. International Journal of
  Forecasting (forthcoming).

\bibitem[{Li and Lim(2018)}]{li2018}
Li, C., Lim, A., 2018. A greedy aggregation-decomposition method for
  intermittent demand forecasting in fashion retailing. European Journal of
  Operational Research 269~(860-869).

\bibitem[{McCabe and Martin(2005)}]{mccabe2005bayesian}
McCabe, B. P.~M., Martin, G.~M., 2005. Bayesian predictions of low count time
  series. International Journal of Forecasting 21, 315--330.

\bibitem[{Prado and West(2010)}]{Prado2010}
Prado, R., West, M., 2010. Time Series: Modelling, Computation \& Inference.
  Chapman \& Hall/CRC Press.

\bibitem[{Seaman(2018)}]{seaman2018}
Seaman, B., 2018. Considerations of a retail forecasting practitioner.
  International Journal of Forecasting (forthcoming).

\bibitem[{Snyder et~al.(2012)Snyder, Ord, and Beaumont}]{snyder2012forecasting}
Snyder, R.~D., Ord, J.~K., Beaumont, A., 2012. Forecasting the intermittent
  demand for slow-moving inventories: A modelling approach. International
  Journal of Forecasting 28, 485--496.

\bibitem[{Syntetos and Boylan(2005)}]{SyntetosBoylan2005}
Syntetos, A.~A., Boylan, J.~E., 2005. The accuracy of intermittent demand
  estimates. International Journal of Forecasting 21~(2), 303--314.

\bibitem[{Terui and Ban(2014)}]{Terui2014}
Terui, N., Ban, M., 2014. Multivariate time series model with hierarchical
  structure for over-dispersed discrete outcomes. Journal of Forecasting 33,
  379--390.

\bibitem[{Teunter and Duncan(2009)}]{Teunter2009}
Teunter, R., Duncan, L., 2009. Forecasting intermittent demand: a comparative
  study. Journal of the Operational Research Society 60, 321--329.

\bibitem[{West and Harrison(1997)}]{west1997book}
West, M., Harrison, J., 1997. Bayesian {F}orecasting and {D}ynamic {M}odels,
  2nd Edition. Springer-Verlag, New York, Inc.

\bibitem[{West et~al.(1985)West, Harrison, and Migon}]{west1985dynamic}
West, M., Harrison, P.~J., Migon, H.~S., 1985. Dynamic generalized linear
  models and {B}ayesian forecasting (with discussion). Journal of the American
  Statistical Association 80~(389), 73--83.

\bibitem[{Willemain et~al.(2004)Willemain, Smart, and Schwarz}]{willemain2004}
Willemain, T.~R., Smart, C.~N., Schwarz, H.~F., 2004. A new approach to
  forecasting intermittent demand for service parts inventories. International
  Journal of Forecasting 20, 375--387.

\bibitem[{Yelland(2009)}]{yelland2009bayesian}
Yelland, P.~M., 2009. Bayesian forecasting for low-count time series using
  state-space models: An empirical evaluation for inventory management.
  International Journal of Production Economics 118, 95--103.

\end{thebibliography}

\end{document}